\documentclass[%
 reprint,
superscriptaddress,
 amsmath,amssymb,
 aps,
]{revtex4-2}

\newcommand{\vncb}{V$_{\rm N}$C$_{\rm B}$}
\newcommand{\cncb}{C$_{\rm N}$C$_{\rm B}$}
\newcommand{\cbon}{C$_{\rm B}$O$_{\rm N}$}
\newcommand{\bn}{B$_{\rm N}$}
\newcommand{\cb}{C$_{\rm B}$}
\newcommand{\on}{O$_{\rm N}$}
\newcommand{\cbcnn}{C$_{\rm B}$C$_{\rm N2}$}
\newcommand{\cncbbb}{C$_{\rm N}$C$_{\rm B3}$}
\usepackage[T1]{fontenc}
\usepackage{graphicx}
\usepackage{dcolumn}
\usepackage{bm}
\usepackage[colorlinks,linkcolor=blue,anchorcolor=blue,citecolor=blue,]{hyperref}

\begin{document}

\preprint{APS/123-QED}

\title{Stark Shift from Quantum Defects in Hexagonal Boron Nitride}

\author{Pei Li}%
\affiliation{Beijing Computational Science Research Center, Beijing 100193, China
}%
\affiliation{School of Integrated Circuit Science and Engineering, Tianjin University of Technology, Tianjin 300384, China}
\author{Ran Xu}%
\affiliation{Beijing Computational Science Research Center, Beijing 100193, China
}%
\affiliation{Université Paris-Saclay, CentraleSupélec, CNRS, Laboratoire SPMS, 91190 Gif-sur-Yvette, France}%
\author{Bing Huang}
\affiliation{Beijing Computational Science Research Center, Beijing 100193, China
}%
\affiliation{School of Physics and Astronomy, Beijing Normal University, Beijing 100875, China}
\author{Song Li}%
 \email{li.song@csrc.ac.cn}
\affiliation{Beijing Computational Science Research Center, Beijing 100193, China
}%

\date{\today}

\begin{abstract}
Color centers in hexagonal boron nitride have emerged as promising candidates for quantum information applications, owing to their efficient and bright single photon emission. Despite the challenges in directly characterizing these emitters, the interaction between external fields and defects, such as the Stark shift, offers valuable insights into their local geometric configurations.
In this study, we focus on clarifying the possible origin of the distinct Stark shift characteristics observed experimentally, particularly in the emission range around 2 eV. We find that the local symmetry of the defects plays a crucial role in determining the nature of the Stark shift, which can be either linear or quadratic. Additionally, the local dielectric environment significantly influences the Stark shift response. Our calculations not only enhance the understanding of the micro-structure of these hitherto unknown emitters but also pave the way for their more effective utilization as single-photon sources and qubits in quantum technologies.    
\end{abstract}

\maketitle


\section{\label{sec:level1}Introduction} 
Optically active defects in hexagonal boron nitride (hBN) have been extensively reported as ideal solid-state single-photon emitters (SPEs) operating at room temperature~\cite{zhang2020material, wolfowicz2021quantum,
sajid2020single,montblanch2023layered,tran2016quantum,tran2016robust}.  The wide optical band gap of hBN (around 6 eV) could host bound states induced by defects without interference from band edge. Therefore, the observed SPEs exhibit high brightness and robustness together with sharp zero-phonon-lines (ZPL) and short lifetime~\cite{tran2016robust}. Moreover, the weak van der Waals interaction among hBN layers offers controllable fabrication and also enables more versatile photonic integration. These outstanding properties facilitate researchers to achieve high quantum collection efficiency, optical imaging~\cite{hayee2020revealing,kianinia2018all}, and deterministic formation towards SPEs in hBN~\cite{xu2021creating}.

The recorded multicolor SPEs at 2 eV in experiment demonstrate spectrum diffusion (ZPL shift) and inhomogeneous broadening which are attributed to variation of local strain and dielectric environment~\cite{tran2016robust}. These phenomenon not only indicates SPEs are sensitive to ambient environment but also shows the possibility to improve the indistinguishability of the photons, the prerequisite of quantum communication, through external tuning methods, e. g. strain~\cite{grosso2017tunable,li2020giant}, temperature~\cite{jungwirth2016temperature}, magnetic and electric field~\cite{noh2018stark,xia2019room,nikolay2019very}. However, the response from emitters varies upon external tuning since the signals come from different defects and their microscopic structures are still unknown. Although optical and electron microscopy could yield information about the defects on top layers of hBN sample~\cite{jin2009fabrication,krivanek2010atom}, experimental identification of the responsible emitters is still challenging since there is no point-to-point connection established among different characterization techniques. At the same time, the emitters might be buried inside deep region that beyond the spatial resolution rather than at surface. With first-principles calculations, several defects are proposed as possible candidate for the observed SPEs, including native defects, carbon \cite{tawfik2017first, sajid2018defect, li2022bistable} and oxygen ~\cite{noh2018stark, li2022identification} impurities. Especially, substitutional carbon defects are believed to highly involved during the formation of the color centers and optically detected magnetic resonance (ODMR) centers~\cite{jara2021first, mendelson2021identifying, chejanovsky2021single}.

Stark tuning has been demonstrated in hBN to minimize spectral diffusion or to match cavity's resonance~\cite{noh2018stark,xia2019room,nikolay2019very}. However, no systematically correspondence has been reported between experimental observation and exact defect structures. Here we theoretically study the Stark shift induced by external electric field perpendicular to hBN layers based on the possible defect models. The local symmetry, especically the out-of-plane distortion at both ground and excited state, determine the shape of Stark shift with tuned electric field. Centrosymmetric defects with $D_{3h}$ symmetry could yield strong quadratic Stark shift. Noncentrosymmetric defects with $C_{s}$ symmetry with out-of-plane distortion at ground and excited state exhibits linear Stark shift, whereas defects with $C_{2v}$ display quasi-quadratic behavior. Our result make an essential step towards the identification of quantum emitters in hBN as well as the great potential of spectral tuning via electric field.

\section{\label{sec:level1}Methods} 
In this Letter, using the density functional theory (DFT) calculations implemented in the \textit{Vienna ab initio simulation package} (VASP) code~\cite{kresse1996efficiency, kresse1996efficient}, we study the Stark shift response from defects in hBN. The cutoff energy for plane wave basis set is 450~eV. The projector augmented wave (PAW) potentials~\cite{blochl1994projector, kresse1999ultrasoft} is used to describe interaction between valence electrons and the core part~\cite{blochl1994projector, kresse1999ultrasoft}. It is noteworthy that the inversion symmetry of hBN should be kept so a slab model with odd number of layers with defect embedded in central layer is necessary. A $6\times6$ trilayer supercell slab is used to mimic the real multilayer system and this is sufficient to use the $\Gamma$-point sampling scheme. The interlayer vdW interaction is described with DFT-D3 method of Grimme~\cite{grimme2010consistent}. To apply electric field, a 18 \AA\ vacuum layer is added in the direction perpendicular to hBN layers. With mixing parameter $\alpha = 0.32$, the hybrid density functional of Heyd, Scuseria, and Ernzerhof (HSE)~\cite{heyd2003hybrid} could yield experimental optical gap around 6~eV without electron-phonon renormalization energy included~\cite{cassabois2016hexagonal} and this functional is used to calculate electronic energy diagram at ground state. The convergence threshold for the forces is set to 0.01~eV/\AA. $\Delta$SCF method~\cite{gali2009theory} is used to calculate electronic excited states.

\section{\label{sec:level1}Result and discussion} 
\begin{figure}
    \includegraphics[width=\columnwidth]{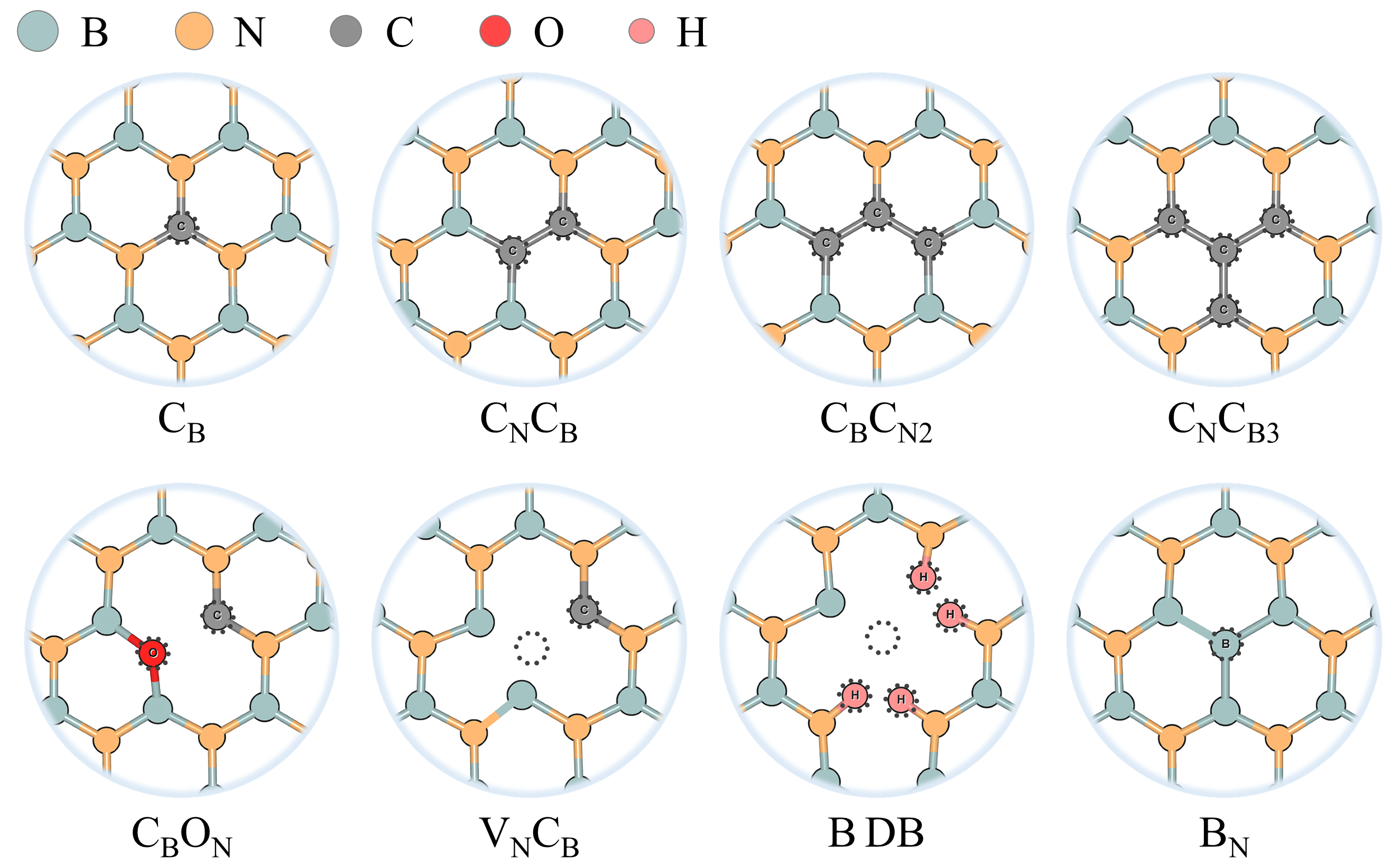}
     \caption{\label{Figure1}%
     Possible defects as single photon source in hBN.}
\end{figure}

The defects we consider here are shown in Fig.~\ref{Figure1} and they all proposed as SPEs either at visible (\cb~\cite{chejanovsky2021single}, \cbcnn~\cite{li2022carbon}, \cncbbb~\cite{benedek2023symmetric}, \cbon~\cite{guo2023coherent}, \vncb~\cite{sajid2018defect,li2022bistable}, B DB~\cite{turiansky2019dangling}), blue (\bn~\cite{li2025native}), or ultraviolet region (\cncb~\cite{mackoit2019carbon}). All the carbon substitutional defects maintain planar configuration without out-of-plane distortion and the simulated phonon side band can match the experiment well. \cb\ and \cncbbb\ are centrosymmetric with $D_{3h}$ symmetry while \cbcnn\ and \cncb\ have $C_{2v}$ symmetry. We proposed \cbon\ previously to explain the narrow ODMR line width observed in experiment. This defect might be challenging to form since both \cb\ and \on\ are donor in hBN and there is large repulsion between them, resulting $C_s$ symmetry. Despite this, the defect remains a subject of study due to its calculated hyperfine constant can well match experimental data~\cite{gao2024single}. The calculated \vncb\ also has out-of-plane configuration at ground state which is consistent with previous result while it retrieves $C_{2v}$ at excited state within our slab model. This is a pseudo Jahn-Teller  system that the excited state is interacted with ground state due to orbital mixing~\cite{li2020giant}. B DB is single boron dangling bond system with all the nitrogen dangling bonds are saturated with hydrogen. We also find the out-of-plane configuration with lower energy in B DB with two upward shifted hydrogen and two downward shifted. Similarly, the \bn\ substitution is also pushed out-of-plane with $C_{3v}$ due to the repulsion from nearby boron atoms.

\begin{figure}
    \includegraphics[width=\columnwidth]{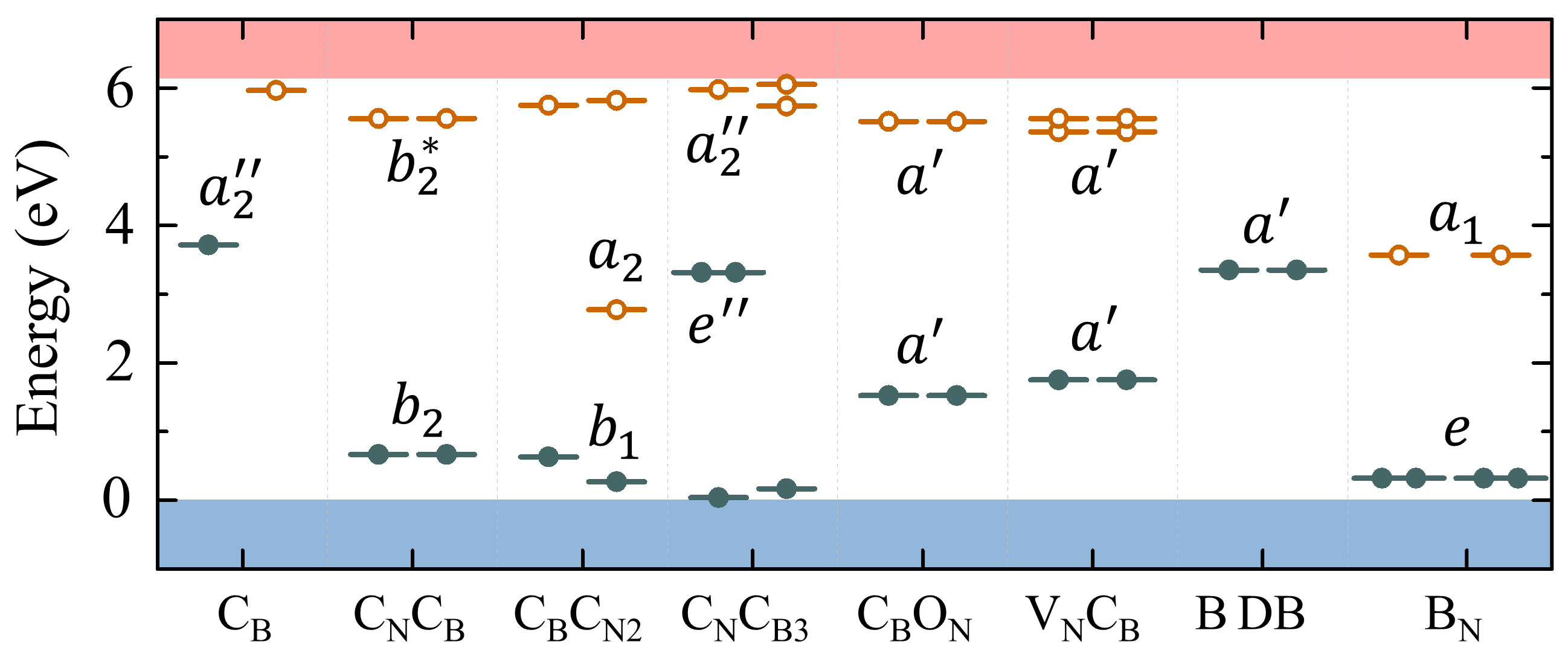}
     \caption{\label{Figure2}%
     The energy diagram of defects we consider. The spin-up and down channel are separated. The occupied and unoccupied localized states are colored with blue and red. }
\end{figure}

\begin{figure}
    \includegraphics[width=\columnwidth]{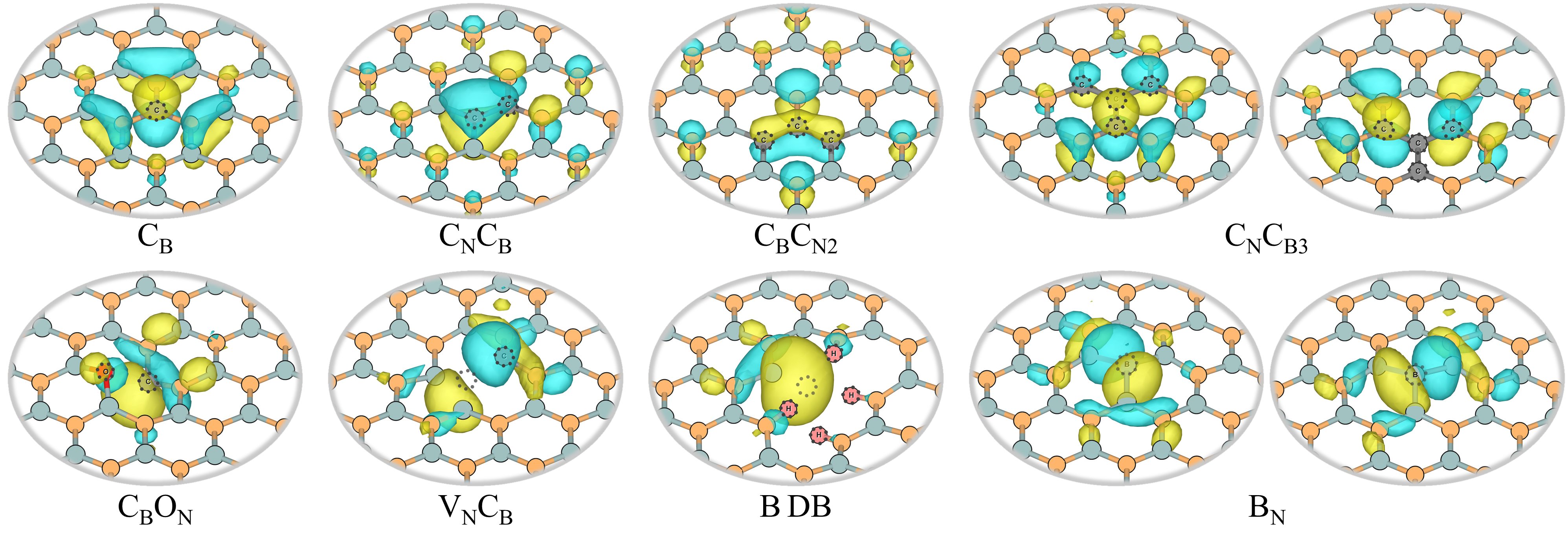}
     \caption{\label{Figure3}%
     The localized wavefunction of defect states that are responsible for optical transition. }
\end{figure}

The energy level diagrams of defects considered are shown in Fig.~\ref{Figure2} with separated spin-up and down channels. The position of several defect levels are slightly shifted compared to previous studies in bulk system, attributed to the quantum confinement in slab model. Nevertheless, the possible optical transition would not change. \cb\ is doublet with one occupied $a^{\prime\prime}_2$ state in gap. \cncb\ is singlet with fully occupied and unoccupied $b_2$ states. This defect has been studied extensively due to its ultraviolet emission at 4.1 eV, however, no Stark shift measurement has been done towards ultraviolet emission. The optical excitation we consider in \cbcnn\ is the lowest transition from $b_1$ to $a_2$ in spin-down channel. \cncbbb\ had transition from degenerate state $e^{\prime\prime}$ to $a^{\prime\prime}_2$, however the dynamic JT effect is not included to fix the symmetry. All these carbon substitution defects are in neutral charge state and the localized defect levels have out-of-plane wavefunction, as shown in Fig.~\ref{Figure3}. The lower $C_s$ symmetry transforms the irreducible representation $a_1$ and $b_1$ in $C_{2v}$ to $a^{\prime}$ in \cbon, \vncb, and B DB. Basically the occupied $a^{\prime}$ is from in-plane orbital while the unoccupied one from out-of-plane orbital. In addition, the B DB is negatively charged and this charge state yields suitable ZPL. To simulate the charge state, oxygen donor \on\ is introduced to avoid the spurious electrostatic interactions that diverge with increased vacuum region~\cite{chou2017nitrogen}. \bn\ is singlet and has occupied $e$ state and unoccupied $a_1$.

\begin{figure}
    \includegraphics[width=\columnwidth]{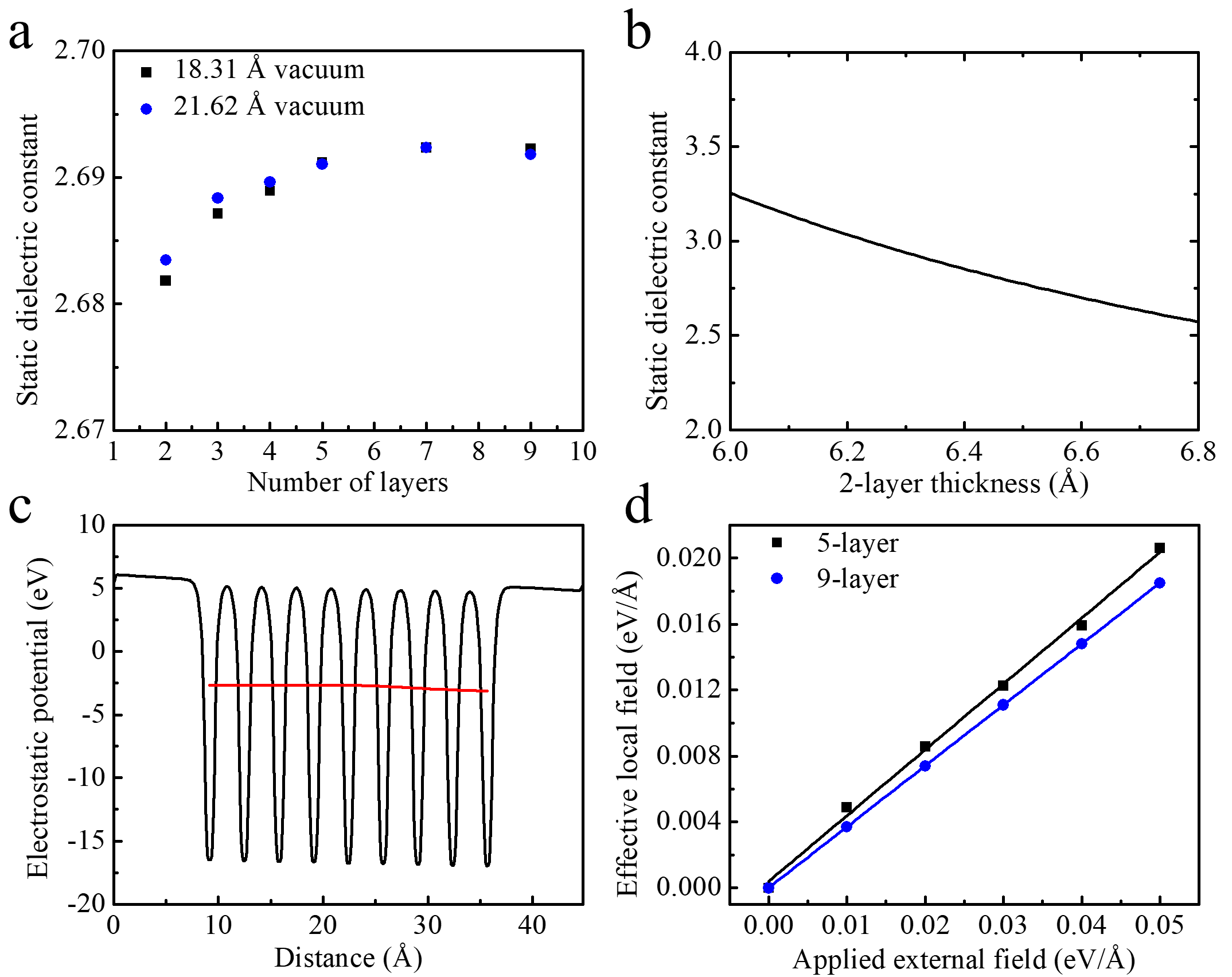}
     \caption{\label{Figure4}%
     The dielectric property of slab model. (a) The layer-dependent out-of-plane dielectric constant with slab containing different vacuum region. (b) The thickness-dependent out-of-plane dielectric constant based on Eq.~\ref{eq1}. (c) Electrostatic potential of 9-layer hBN with applied electric field 0.05 eV/\AA. The red line is the moving average of electrostatic potential inside hBN. The fitted slope of moving average is the effective local field which is 0.0185 eV/\AA. (d) The fitted effective local field with various applied electric field. The slope of linear fitting is close to $\epsilon_{h,\perp}^{-1}$. }
\end{figure}

The hBN responds to external electric field through charge redistribution. This generates dipoles inside the layers and the electric field can be described in terms of the dielectric constant. To verify the convergence of the trilayer model, we calculate the out-of-plane dielectric constant via following equation:
\begin{equation}\label{eq1}
\epsilon_{h,\perp} = [1 + \frac{s}{h}(\frac{1}{\epsilon_{s,\perp}} - 1)]^{-1},
\end{equation}
where $\epsilon_{s,\perp}$ and $\epsilon_{h,\perp}$ represent the out-of-plane dielectric constant of the supercell and slab. To eliminate the contribution of vacuum layer, the $\epsilon_{h,\perp}$ should be rescaled based on the thickness of supercell $s$ and slab $h$. Single layer thickness is defined with the van der Waals distance. The calculated $\epsilon_{h,\perp}$ in Fig.~\ref{Figure4}a is around 2.69 showing relative small thickness dependence which in contrast to previous study due to the fixed thickness 3.31 \AA\ is used here~\cite{laturia2018dielectric}. As shown in Fig.~\ref{Figure4}b, the dielectric constant can be simply rescaled with various thickness values and it could be up to 3.25. Due to the dielectric screening, the electric field that defects experienced is different from the field applied on the slab model with vacuum. The effective local field can be calculated with moving average of the electrostatic potential inside the slab~\cite{alaerts2024first,bathen2020first}. To analyze the relationship between the averaged effective local field and various applied fields, we employed 5- and 9-layer slab models, as the 3-layer slab was insufficient to yield a smooth slope. The slope ultimately converges to $\epsilon_{h,\perp}^{-1}$. Therefore the corresponding macroscopic electric field $E_z$ is range from -1.85 to 1.85 $MV/cm$ which can be easily achieved in experiment.

With $E_z$ on defect, the Stark shift of the ZPL can be given by:
\begin{equation}\label{eq2}
\Delta_{ZPL} = -\Delta{\mu_{z}}\cdot E_z-\frac{1}{2}E_z\cdot \Delta\alpha_z\cdot E_z,
\end{equation}
where $\Delta{\mu}$ and $\Delta\alpha$ are the relative change in electric dipole moment and polarizability along $z$ component between ground and excited states. Also, these two parameters correspond to the linear and quadratic coefficients. High order hyperpolarizability rarely appear in low electric field and has not been reported in hBN while it indeed has been observed in SnV in diamond at high field where third even fourth-order term should be included~\cite{de2021investigation}. To evaluate the two parameters, we calculate the $\Delta_{ZPL}$ from ground to excited states under a set of electric fields.

\begin{figure}
    \includegraphics[width=\columnwidth]{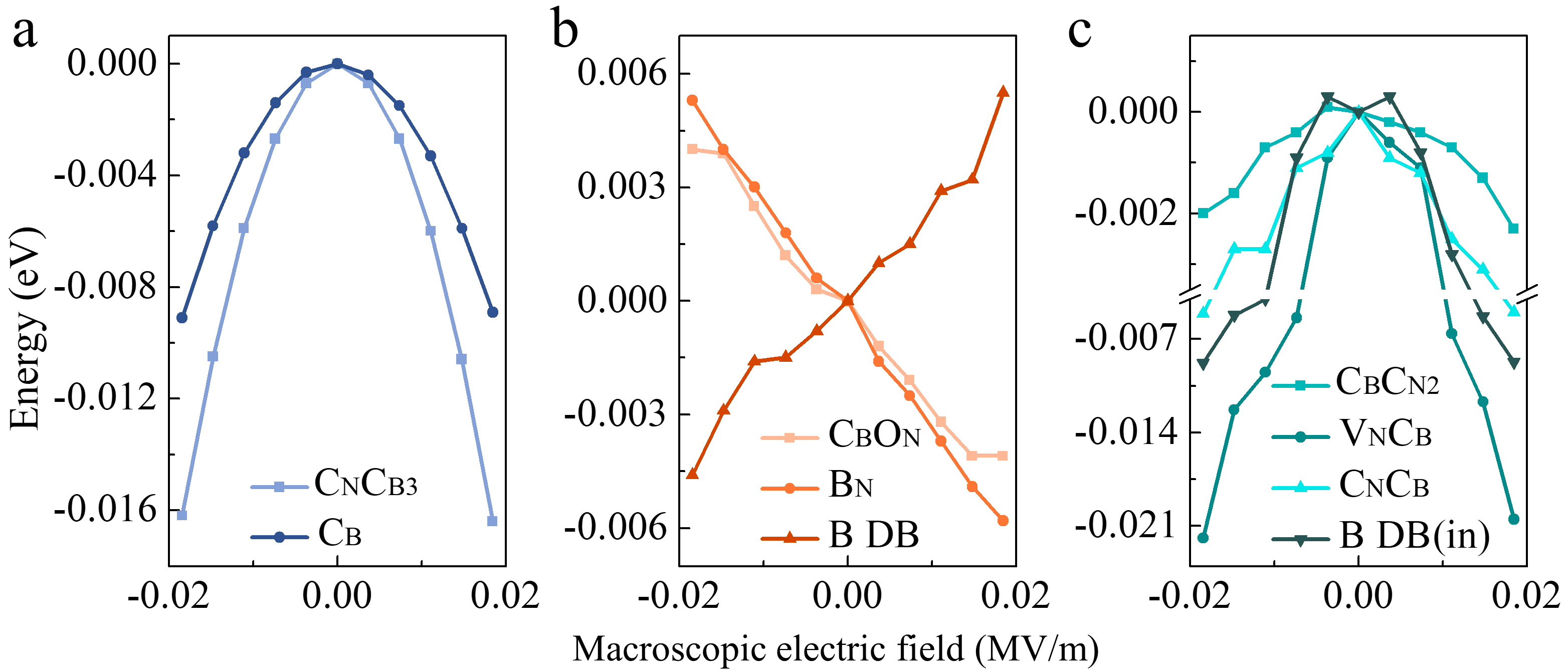}
     \caption{\label{Figure5}%
     The Stark shift of defects with various symmetry. The energy shift is aligned to ZPL without electric field. }
\end{figure}

Fig.~\ref{Figure5} demonstrate the Stark behaviors of defects consider here and we classify them into three groups. (1) centrosymmetric defects with quadratic Stark shifts due to $D_{3h}$ symmetry; (2) defects with out-of-plane distortions, exhibiting strong linear Stark shifts due to broken inversion symmetry; and (3) defects with quasi-quadratic Stark shifts, typically showing $C_{2v}$ symmetry. Centrosymmetric defects \cb\ and \cncbbb\ with $D_{3h}$ symmetry yield strong quadratic Stark shift. With above equation, the fitted $\Delta\alpha$ are 53 $\AA^3$ and 95.6 $\AA^3$, respectively, for \cb\ and \cncbbb. The polarizability is proportional to the polarizability volume and the smaller polarizability of \cb\ indicates the small space that the polarization extends and this can also be reflected by the more localized wavefunction in Fig.~\ref{Figure3}. The positive transition polarizability indicate the stronger polarizability of excited state than ground state, as shown in Supplementary Figure 1. Interestingly, we also note the relative displacement of the atoms under increased electric field, or the piezoelectric effect, acting similar to strain~\cite{maze2011properties,udvarhelyi2023planar}. This can be explained by changes in electrostatic potential close to the defect under electric field. The local charge redistribution generates quantum mechanical forces on the atoms. The decreased relaxation energy of excited state with increased electric field indicates the possibility to modulate the ZPL line and photoluminescence intensity with high electric field (see Supplementary Figure 2). 

In Fig. ~\ref{Figure5}b, the defects with out-of-plane distortion at ground and excited state could break the inversion symmetry and induce permanent dipole in the out-of-plane direction and show strong linear Stark shift with $\Delta{\mu}$ to be 1.15 $D$ in \cbon, 1.44 $D$ in \bn, and -1.14 $D$ in B DB. The sign of the value corresponds to the two mirror symmetric out-of-plane distortion. This result is consistent with experiment value from -0.9 to 0.9 $D$~\cite{noh2018stark}. We speculate the larger out-of-plane distortion could yield higher $\Delta{\mu}$ and 30 meV/(V/nm) change in \bn\ matches the giant Stark shift reported previously~\cite{xia2019room}. One more case with carbon tetramer with no linear Stark shift is shown in Supplementary Figure 3. The planar configuration with mirror symmetry makes it lack of out-of-plane dipole. 

We plot the quasi-quadratic Stark shift in Fig.~\ref{Figure5}c. Generally all the defects here have $C_{2v}$ symmetry either at ground or at excited state. Although the inversion symmetry is kept, the lack of rotation symmetry makes the polarizability in carbon substitution smaller. $\Delta\alpha$ is 12.8 $\AA^3$ and 26 $\AA^3$ for \cncb\ and \cbcnn, respectively. We also show the B DB defect with in-plane configuration as comparison to address that the mirror symmetry determines whether the Stark shift is linear or quadratic. The fitted $\Delta{\mu}$ $=$ 0.05 $D$ with $\Delta\alpha$ $=$ 53.7 $\AA^3$. In experiment, the existence of in-plane strain can cause the variation of polarizability and dipole orientation, and then change the quadratic shift and the out-of-plane might alter the quadratic shift to linear. Another phenomenon is the misalignment between absorption and emission dipole orientation observed previously~\cite{noh2018stark}. This can be explained by the resonant excitation to the conduction band minimum (CBM) or states close to CBM and then polarization loss. For example, the CBM is heavily involved during excitation in \cb\ and B DB while the electrons have chance to be promoted to CBM with higher energy in \cncbbb\ since the empty states are just below CBM.

As mentioned above, here we find the out-of-plane distortion in \vncb\ ground state, however the optimized excited state retrieves the planar configuration with lower energy. This geometry change during excitation significantly modifies the linear shape into V-shape. The fitted $\Delta{\mu}$ $=$ 0.32 $D$ which is close to experimental date 0.24 $D$ with $\Delta\alpha$ $=$ 121.8 $\AA^3$ that is missing in previous study~\cite{noh2018stark}. We should note this kind of geometry change in \vncb\ usually yields larger Huang-Rhys factor and phonon side band, resulting undistinguished ZPL and Stark shift. Therefore the potential defect structures for V-shape should be one with tiny relaxation during excitation from ground to excited state while symmetry breaking still exists during the excitation. Another possible explanation is that the defects should strongly couple with electric field that even small electric field could change the direction of the electric dipole moment. The out-of-plane breathing mode or dynamic JT effect might contribute to this. The external magnetic field is a effective to flip the sign of dipole moment for exciton emitter, nevertheless its might be not the case in defect systems since the electron and hole are usually localized.

We would like to mention the Stark shift regarding the number of layers of the slab. When it comes to the intra-defect optical transition, e. g. \bn, the ZPL shows negligible variation while the ZPL changes when the band edge is involved, for example in \cb, as depicted in Supplementary Figure 4. This is mainly due to the quantum confinement that the band gap increases as the decreasing of number of layers. Therefore the fitted polarizability $\Delta\alpha$ of \cb\ demonstrates layer-dependence and increases to 256 $\AA^3$. This is a computational issue that the slab thickness and vacuum size could influence the result. However, this further remind us in experiment the local environment, for example the different polytypes of BN~\cite{iwanski2024revealing,ohba2001first}, might also modify the Stark shift response since the band gap differs. Another issue is the definition of thickness of slab mentioned above which can strongly modify the calculated dielectric constant as shown in Fig.~\ref{Figure4}b based on Eq.~\ref{eq1}. Earlier works defined the thickness either based on the total extent of the polarization~\cite{kumar2016thickness} or the van der Waals bond length~\cite{laturia2018dielectric}. Both are distinct from and smaller than the physical interlayer distance. Once we adopt $\epsilon_{h,\perp}$ = 3.52, the calculated Stark shift parameters can be readjusted as listed in Tab.~\ref{tab:data}. This might be another possible reason for the huge difference exists among the measured polarizability from emitters besides the defect geometry difference~\cite{noh2018stark,nikolay2019very,zhigulin2023stark}. With larger dielectric constant,  the polarizability of \cncbbb\ is very close to previous experiment result around 150 $\AA^3$ ~\cite{noh2018stark} whereas the simulated $\Delta{\mu}$ is only 0.02 $D$ which is one order smaller (0.22 $D$). The ZPL of \cncbbb\ from HSE calculation is 2.04 eV which is not far from 1.88 eV (658.4 nm).

\renewcommand{\arraystretch}{1.5}
\begin{table}[htb]
\caption{\label{tab:data} Calculated dipole moment and polarizability based on different dielectric constant $\epsilon_{h,\perp}$. B DB (in) indicates the B DB defect with $C_{2v}$ symmetry.}
\begin{ruledtabular}
\begin{tabular}{l|cc}
Defect & $\epsilon_{h,\perp}$ = 2.69 & $\epsilon_{h,\perp}$ = 3.52 \\
\hline
\cb  & 53.0 $\AA^3$ & 98.5 $\AA^3$\\
\cncbbb  & 96.5 $\AA^3$ & 161.3 $\AA^3$\\
\cbon & 1.15 $D$ & 1.49 $D$\\
\bn & 1.44 $D$ & 1.87 $D$\\
B DB & -1.14 $D$ & -1.49 $D$\\
\cncb & 12.8 $\AA^3$ & 21.6 $\AA^3$\\
\cbcnn & 26.0 $\AA^3$ & 43.9 $\AA^3$\\
B DB (in) & 53.7 $\AA^3$ & 90.6 $\AA^3$\\
\end{tabular}
\end{ruledtabular}
\end{table}

An alternative method to calculate the dipole moment is based on modern theory of polarization via Berry phase expressions~\cite{king1993theory,resta1994macroscopic}. Although this method could be directly applied to bulk system, it is only suitable for defects with possible centrosymmetric configuration as reference so that the polarization changes could be calculated. More detailed discussion can be found in Ref.~\cite{alaerts2024first,spaldin2012beginner}. Previous studies regarding NV center in diamond show discrepancy between the calculated dipole moment from the fitting mentioned above based on electric field and the modern theory of polarization~\cite{udvarhelyi2019spectrally,alaerts2024first}. To test this on hBN, we select \bn\ as an example since the planar configuration has $D_{3h}$ symmetry. The calculated dipole moment is 4.92 $D$. We notice that the modern theory implemented in VASP might be not compatible with the $\Delta$SCF method that fixing the occupation of the electrons. To further optimize this, we project the dipole moment component to specified defect levels inside band gap. So the dipole moment change during excitation can be derived from single defect level. With this method, we obtain $\Delta{\mu}$ $=$ 2.91 $D$ which is still overestimated.

In summary, our calculations provide valuable information towards the SPEs identification in hBN. Based on the Stark shift behaviors, we link the experimentally observed data with exact defect structures including common impurities e. g. carbon, oxygen and hydrogen. \cncbbb\ shows quadratic Stark shift with polarizability close to previous measurement. The out-of-plane distortion in either ground or excited state of the defect determines the shape of Stark shift. The mirror symmetry breaking is essential to have out-of-plane dipole moment and linear Stark shift. \cbon, \bn\ and B DB could have transition dipole moment that are consistent with experiment. We tentatively associate the V-shape Stark shift with JT system or defects strongly coupled with electric field. We also address the local dielectric environment could remarkably modify the Stark shift. In other words, once we know the exact structure of quantum defect, it can be used as highly sensitive quantum sensor to probe local dielectric environment.

\begin{acknowledgments}
BH acknowledges the National Key Research and Development of China (Grant No.\ 2022YFA1402400), NSFC (Grant No.\ 12088101), and NSAF (Grant No.\ U2230402). PL acknowledges the NSFC (Grant No.\ 12404094). This research was supported by Open Research Fund of CNMGE Platform $\&$ NSCC-TJ.
\end{acknowledgments}

\bibliography{apssamp}

\begin{thebibliography}{54}%
\makeatletter
\providecommand \@ifxundefined [1]{%
 \@ifx{#1\undefined}
}%
\providecommand \@ifnum [1]{%
 \ifnum #1\expandafter \@firstoftwo
 \else \expandafter \@secondoftwo
 \fi
}%
\providecommand \@ifx [1]{%
 \ifx #1\expandafter \@firstoftwo
 \else \expandafter \@secondoftwo
 \fi
}%
\providecommand \natexlab [1]{#1}%
\providecommand \enquote  [1]{``#1''}%
\providecommand \bibnamefont  [1]{#1}%
\providecommand \bibfnamefont [1]{#1}%
\providecommand \citenamefont [1]{#1}%
\providecommand \href@noop [0]{\@secondoftwo}%
\providecommand \href [0]{\begingroup \@sanitize@url \@href}%
\providecommand \@href[1]{\@@startlink{#1}\@@href}%
\providecommand \@@href[1]{\endgroup#1\@@endlink}%
\providecommand \@sanitize@url [0]{\catcode `\\12\catcode `\$12\catcode `\&12\catcode `\#12\catcode `\^12\catcode `\_12\catcode `\%12\relax}%
\providecommand \@@startlink[1]{}%
\providecommand \@@endlink[0]{}%
\providecommand \url  [0]{\begingroup\@sanitize@url \@url }%
\providecommand \@url [1]{\endgroup\@href {#1}{\urlprefix }}%
\providecommand \urlprefix  [0]{URL }%
\providecommand \Eprint [0]{\href }%
\providecommand \doibase [0]{https://doi.org/}%
\providecommand \selectlanguage [0]{\@gobble}%
\providecommand \bibinfo  [0]{\@secondoftwo}%
\providecommand \bibfield  [0]{\@secondoftwo}%
\providecommand \translation [1]{[#1]}%
\providecommand \BibitemOpen [0]{}%
\providecommand \bibitemStop [0]{}%
\providecommand \bibitemNoStop [0]{.\EOS\space}%
\providecommand \EOS [0]{\spacefactor3000\relax}%
\providecommand \BibitemShut  [1]{\csname bibitem#1\endcsname}%
\let\auto@bib@innerbib\@empty
\bibitem [{\citenamefont {Zhang}\ \emph {et~al.}(2020)\citenamefont {Zhang}, \citenamefont {Cheng}, \citenamefont {Chou},\ and\ \citenamefont {Gali}}]{zhang2020material}%
  \BibitemOpen
  \bibfield  {author} {\bibinfo {author} {\bibfnamefont {G.}~\bibnamefont {Zhang}}, \bibinfo {author} {\bibfnamefont {Y.}~\bibnamefont {Cheng}}, \bibinfo {author} {\bibfnamefont {J.-P.}\ \bibnamefont {Chou}},\ and\ \bibinfo {author} {\bibfnamefont {A.}~\bibnamefont {Gali}},\ }\bibfield  {title} {\bibinfo {title} {Material platforms for defect qubits and single-photon emitters},\ }\href@noop {} {\bibfield  {journal} {\bibinfo  {journal} {Appl. Phys. Rev.}\ }\textbf {\bibinfo {volume} {7}},\ \bibinfo {pages} {031308} (\bibinfo {year} {2020})}\BibitemShut {NoStop}%
\bibitem [{\citenamefont {Wolfowicz}\ \emph {et~al.}(2021)\citenamefont {Wolfowicz}, \citenamefont {Heremans}, \citenamefont {Anderson}, \citenamefont {Kanai}, \citenamefont {Seo}, \citenamefont {Gali}, \citenamefont {Galli},\ and\ \citenamefont {Awschalom}}]{wolfowicz2021quantum}%
  \BibitemOpen
  \bibfield  {author} {\bibinfo {author} {\bibfnamefont {G.}~\bibnamefont {Wolfowicz}}, \bibinfo {author} {\bibfnamefont {F.~J.}\ \bibnamefont {Heremans}}, \bibinfo {author} {\bibfnamefont {C.~P.}\ \bibnamefont {Anderson}}, \bibinfo {author} {\bibfnamefont {S.}~\bibnamefont {Kanai}}, \bibinfo {author} {\bibfnamefont {H.}~\bibnamefont {Seo}}, \bibinfo {author} {\bibfnamefont {A.}~\bibnamefont {Gali}}, \bibinfo {author} {\bibfnamefont {G.}~\bibnamefont {Galli}},\ and\ \bibinfo {author} {\bibfnamefont {D.~D.}\ \bibnamefont {Awschalom}},\ }\bibfield  {title} {\bibinfo {title} {Quantum guidelines for solid-state spin defects},\ }\href@noop {} {\bibfield  {journal} {\bibinfo  {journal} {Nat. Rev. Mater.}\ }\textbf {\bibinfo {volume} {6}},\ \bibinfo {pages} {906} (\bibinfo {year} {2021})}\BibitemShut {NoStop}%
\bibitem [{\citenamefont {Sajid}\ \emph {et~al.}(2020)\citenamefont {Sajid}, \citenamefont {Ford},\ and\ \citenamefont {Reimers}}]{sajid2020single}%
  \BibitemOpen
  \bibfield  {author} {\bibinfo {author} {\bibfnamefont {A.}~\bibnamefont {Sajid}}, \bibinfo {author} {\bibfnamefont {M.~J.}\ \bibnamefont {Ford}},\ and\ \bibinfo {author} {\bibfnamefont {J.~R.}\ \bibnamefont {Reimers}},\ }\bibfield  {title} {\bibinfo {title} {Single-photon emitters in hexagonal boron nitride: a review of progress},\ }\href@noop {} {\bibfield  {journal} {\bibinfo  {journal} {Rep. Prog. Phys.}\ }\textbf {\bibinfo {volume} {83}},\ \bibinfo {pages} {044501} (\bibinfo {year} {2020})}\BibitemShut {NoStop}%
\bibitem [{\citenamefont {Montblanch}\ \emph {et~al.}(2023)\citenamefont {Montblanch}, \citenamefont {Barbone}, \citenamefont {Aharonovich}, \citenamefont {Atat{\"u}re},\ and\ \citenamefont {Ferrari}}]{montblanch2023layered}%
  \BibitemOpen
  \bibfield  {author} {\bibinfo {author} {\bibfnamefont {A.~R.-P.}\ \bibnamefont {Montblanch}}, \bibinfo {author} {\bibfnamefont {M.}~\bibnamefont {Barbone}}, \bibinfo {author} {\bibfnamefont {I.}~\bibnamefont {Aharonovich}}, \bibinfo {author} {\bibfnamefont {M.}~\bibnamefont {Atat{\"u}re}},\ and\ \bibinfo {author} {\bibfnamefont {A.~C.}\ \bibnamefont {Ferrari}},\ }\bibfield  {title} {\bibinfo {title} {Layered materials as a platform for quantum technologies},\ }\href@noop {} {\bibfield  {journal} {\bibinfo  {journal} {Nat. Nanotechnol.}\ }\textbf {\bibinfo {volume} {18}},\ \bibinfo {pages} {555} (\bibinfo {year} {2023})}\BibitemShut {NoStop}%
\bibitem [{\citenamefont {Tran}\ \emph {et~al.}(2016{\natexlab{a}})\citenamefont {Tran}, \citenamefont {Bray}, \citenamefont {Ford}, \citenamefont {Toth},\ and\ \citenamefont {Aharonovich}}]{tran2016quantum}%
  \BibitemOpen
  \bibfield  {author} {\bibinfo {author} {\bibfnamefont {T.~T.}\ \bibnamefont {Tran}}, \bibinfo {author} {\bibfnamefont {K.}~\bibnamefont {Bray}}, \bibinfo {author} {\bibfnamefont {M.~J.}\ \bibnamefont {Ford}}, \bibinfo {author} {\bibfnamefont {M.}~\bibnamefont {Toth}},\ and\ \bibinfo {author} {\bibfnamefont {I.}~\bibnamefont {Aharonovich}},\ }\bibfield  {title} {\bibinfo {title} {Quantum emission from hexagonal boron nitride monolayers},\ }\href@noop {} {\bibfield  {journal} {\bibinfo  {journal} {Nat. Nanotechnol.}\ }\textbf {\bibinfo {volume} {11}},\ \bibinfo {pages} {37} (\bibinfo {year} {2016}{\natexlab{a}})}\BibitemShut {NoStop}%
\bibitem [{\citenamefont {Tran}\ \emph {et~al.}(2016{\natexlab{b}})\citenamefont {Tran}, \citenamefont {Elbadawi}, \citenamefont {Totonjian}, \citenamefont {Lobo}, \citenamefont {Grosso}, \citenamefont {Moon}, \citenamefont {Englund}, \citenamefont {Ford}, \citenamefont {Aharonovich},\ and\ \citenamefont {Toth}}]{tran2016robust}%
  \BibitemOpen
  \bibfield  {author} {\bibinfo {author} {\bibfnamefont {T.~T.}\ \bibnamefont {Tran}}, \bibinfo {author} {\bibfnamefont {C.}~\bibnamefont {Elbadawi}}, \bibinfo {author} {\bibfnamefont {D.}~\bibnamefont {Totonjian}}, \bibinfo {author} {\bibfnamefont {C.~J.}\ \bibnamefont {Lobo}}, \bibinfo {author} {\bibfnamefont {G.}~\bibnamefont {Grosso}}, \bibinfo {author} {\bibfnamefont {H.}~\bibnamefont {Moon}}, \bibinfo {author} {\bibfnamefont {D.~R.}\ \bibnamefont {Englund}}, \bibinfo {author} {\bibfnamefont {M.~J.}\ \bibnamefont {Ford}}, \bibinfo {author} {\bibfnamefont {I.}~\bibnamefont {Aharonovich}},\ and\ \bibinfo {author} {\bibfnamefont {M.}~\bibnamefont {Toth}},\ }\bibfield  {title} {\bibinfo {title} {Robust multicolor single photon emission from point defects in hexagonal boron nitride},\ }\href@noop {} {\bibfield  {journal} {\bibinfo  {journal} {ACS Nano}\ }\textbf {\bibinfo {volume} {10}},\ \bibinfo {pages} {7331} (\bibinfo {year} {2016}{\natexlab{b}})}\BibitemShut {NoStop}%
\bibitem [{\citenamefont {Hayee}\ \emph {et~al.}(2020)\citenamefont {Hayee}, \citenamefont {Yu}, \citenamefont {Zhang}, \citenamefont {Ciccarino}, \citenamefont {Nguyen}, \citenamefont {Marshall}, \citenamefont {Aharonovich}, \citenamefont {Vu{\v{c}}kovi{\'c}}, \citenamefont {Narang}, \citenamefont {Heinz} \emph {et~al.}}]{hayee2020revealing}%
  \BibitemOpen
  \bibfield  {author} {\bibinfo {author} {\bibfnamefont {F.}~\bibnamefont {Hayee}}, \bibinfo {author} {\bibfnamefont {L.}~\bibnamefont {Yu}}, \bibinfo {author} {\bibfnamefont {J.~L.}\ \bibnamefont {Zhang}}, \bibinfo {author} {\bibfnamefont {C.~J.}\ \bibnamefont {Ciccarino}}, \bibinfo {author} {\bibfnamefont {M.}~\bibnamefont {Nguyen}}, \bibinfo {author} {\bibfnamefont {A.~F.}\ \bibnamefont {Marshall}}, \bibinfo {author} {\bibfnamefont {I.}~\bibnamefont {Aharonovich}}, \bibinfo {author} {\bibfnamefont {J.}~\bibnamefont {Vu{\v{c}}kovi{\'c}}}, \bibinfo {author} {\bibfnamefont {P.}~\bibnamefont {Narang}}, \bibinfo {author} {\bibfnamefont {T.~F.}\ \bibnamefont {Heinz}}, \emph {et~al.},\ }\bibfield  {title} {\bibinfo {title} {Revealing multiple classes of stable quantum emitters in hexagonal boron nitride with correlated optical and electron microscopy},\ }\href@noop {} {\bibfield  {journal} {\bibinfo  {journal} {Nat. Mater.}\ }\textbf {\bibinfo {volume} {19}},\ \bibinfo {pages} {534} (\bibinfo {year}
  {2020})}\BibitemShut {NoStop}%
\bibitem [{\citenamefont {Kianinia}\ \emph {et~al.}(2018)\citenamefont {Kianinia}, \citenamefont {Bradac}, \citenamefont {Sontheimer}, \citenamefont {Wang}, \citenamefont {Tran}, \citenamefont {Nguyen}, \citenamefont {Kim}, \citenamefont {Xu}, \citenamefont {Jin}, \citenamefont {Schell} \emph {et~al.}}]{kianinia2018all}%
  \BibitemOpen
  \bibfield  {author} {\bibinfo {author} {\bibfnamefont {M.}~\bibnamefont {Kianinia}}, \bibinfo {author} {\bibfnamefont {C.}~\bibnamefont {Bradac}}, \bibinfo {author} {\bibfnamefont {B.}~\bibnamefont {Sontheimer}}, \bibinfo {author} {\bibfnamefont {F.}~\bibnamefont {Wang}}, \bibinfo {author} {\bibfnamefont {T.~T.}\ \bibnamefont {Tran}}, \bibinfo {author} {\bibfnamefont {M.}~\bibnamefont {Nguyen}}, \bibinfo {author} {\bibfnamefont {S.}~\bibnamefont {Kim}}, \bibinfo {author} {\bibfnamefont {Z.-Q.}\ \bibnamefont {Xu}}, \bibinfo {author} {\bibfnamefont {D.}~\bibnamefont {Jin}}, \bibinfo {author} {\bibfnamefont {A.~W.}\ \bibnamefont {Schell}}, \emph {et~al.},\ }\bibfield  {title} {\bibinfo {title} {All-optical control and super-resolution imaging of quantum emitters in layered materials},\ }\href@noop {} {\bibfield  {journal} {\bibinfo  {journal} {Nat. Commun.}\ }\textbf {\bibinfo {volume} {9}},\ \bibinfo {pages} {874} (\bibinfo {year} {2018})}\BibitemShut {NoStop}%
\bibitem [{\citenamefont {Xu}\ \emph {et~al.}(2021)\citenamefont {Xu}, \citenamefont {Martin}, \citenamefont {Sychev}, \citenamefont {Lagutchev}, \citenamefont {Chen}, \citenamefont {Taniguchi}, \citenamefont {Watanabe}, \citenamefont {Shalaev},\ and\ \citenamefont {Boltasseva}}]{xu2021creating}%
  \BibitemOpen
  \bibfield  {author} {\bibinfo {author} {\bibfnamefont {X.}~\bibnamefont {Xu}}, \bibinfo {author} {\bibfnamefont {Z.~O.}\ \bibnamefont {Martin}}, \bibinfo {author} {\bibfnamefont {D.}~\bibnamefont {Sychev}}, \bibinfo {author} {\bibfnamefont {A.~S.}\ \bibnamefont {Lagutchev}}, \bibinfo {author} {\bibfnamefont {Y.~P.}\ \bibnamefont {Chen}}, \bibinfo {author} {\bibfnamefont {T.}~\bibnamefont {Taniguchi}}, \bibinfo {author} {\bibfnamefont {K.}~\bibnamefont {Watanabe}}, \bibinfo {author} {\bibfnamefont {V.~M.}\ \bibnamefont {Shalaev}},\ and\ \bibinfo {author} {\bibfnamefont {A.}~\bibnamefont {Boltasseva}},\ }\bibfield  {title} {\bibinfo {title} {Creating quantum emitters in hexagonal boron nitride deterministically on chip-compatible substrates},\ }\href@noop {} {\bibfield  {journal} {\bibinfo  {journal} {Nano Lett.}\ }\textbf {\bibinfo {volume} {21}},\ \bibinfo {pages} {8182} (\bibinfo {year} {2021})}\BibitemShut {NoStop}%
\bibitem [{\citenamefont {Grosso}\ \emph {et~al.}(2017)\citenamefont {Grosso}, \citenamefont {Moon}, \citenamefont {Lienhard}, \citenamefont {Ali}, \citenamefont {Efetov}, \citenamefont {Furchi}, \citenamefont {Jarillo-Herrero}, \citenamefont {Ford}, \citenamefont {Aharonovich},\ and\ \citenamefont {Englund}}]{grosso2017tunable}%
  \BibitemOpen
  \bibfield  {author} {\bibinfo {author} {\bibfnamefont {G.}~\bibnamefont {Grosso}}, \bibinfo {author} {\bibfnamefont {H.}~\bibnamefont {Moon}}, \bibinfo {author} {\bibfnamefont {B.}~\bibnamefont {Lienhard}}, \bibinfo {author} {\bibfnamefont {S.}~\bibnamefont {Ali}}, \bibinfo {author} {\bibfnamefont {D.~K.}\ \bibnamefont {Efetov}}, \bibinfo {author} {\bibfnamefont {M.~M.}\ \bibnamefont {Furchi}}, \bibinfo {author} {\bibfnamefont {P.}~\bibnamefont {Jarillo-Herrero}}, \bibinfo {author} {\bibfnamefont {M.~J.}\ \bibnamefont {Ford}}, \bibinfo {author} {\bibfnamefont {I.}~\bibnamefont {Aharonovich}},\ and\ \bibinfo {author} {\bibfnamefont {D.}~\bibnamefont {Englund}},\ }\bibfield  {title} {\bibinfo {title} {Tunable and high-purity room temperature single-photon emission from atomic defects in hexagonal boron nitride},\ }\href@noop {} {\bibfield  {journal} {\bibinfo  {journal} {Nat. Commun.}\ }\textbf {\bibinfo {volume} {8}},\ \bibinfo {pages} {705} (\bibinfo {year} {2017})}\BibitemShut {NoStop}%
\bibitem [{\citenamefont {Li}\ \emph {et~al.}(2020)\citenamefont {Li}, \citenamefont {Chou}, \citenamefont {Hu}, \citenamefont {Plenio}, \citenamefont {Udvarhelyi}, \citenamefont {Thiering}, \citenamefont {Abdi},\ and\ \citenamefont {Gali}}]{li2020giant}%
  \BibitemOpen
  \bibfield  {author} {\bibinfo {author} {\bibfnamefont {S.}~\bibnamefont {Li}}, \bibinfo {author} {\bibfnamefont {J.-P.}\ \bibnamefont {Chou}}, \bibinfo {author} {\bibfnamefont {A.}~\bibnamefont {Hu}}, \bibinfo {author} {\bibfnamefont {M.~B.}\ \bibnamefont {Plenio}}, \bibinfo {author} {\bibfnamefont {P.}~\bibnamefont {Udvarhelyi}}, \bibinfo {author} {\bibfnamefont {G.}~\bibnamefont {Thiering}}, \bibinfo {author} {\bibfnamefont {M.}~\bibnamefont {Abdi}},\ and\ \bibinfo {author} {\bibfnamefont {A.}~\bibnamefont {Gali}},\ }\bibfield  {title} {\bibinfo {title} {Giant shift upon strain on the fluorescence spectrum of vnnb color centers in h-bn},\ }\href@noop {} {\bibfield  {journal} {\bibinfo  {journal} {npj Quantum Inf.}\ }\textbf {\bibinfo {volume} {6}},\ \bibinfo {pages} {85} (\bibinfo {year} {2020})}\BibitemShut {NoStop}%
\bibitem [{\citenamefont {Jungwirth}\ \emph {et~al.}(2016)\citenamefont {Jungwirth}, \citenamefont {Calderon}, \citenamefont {Ji}, \citenamefont {Spencer}, \citenamefont {Flatt{\'e}},\ and\ \citenamefont {Fuchs}}]{jungwirth2016temperature}%
  \BibitemOpen
  \bibfield  {author} {\bibinfo {author} {\bibfnamefont {N.~R.}\ \bibnamefont {Jungwirth}}, \bibinfo {author} {\bibfnamefont {B.}~\bibnamefont {Calderon}}, \bibinfo {author} {\bibfnamefont {Y.}~\bibnamefont {Ji}}, \bibinfo {author} {\bibfnamefont {M.~G.}\ \bibnamefont {Spencer}}, \bibinfo {author} {\bibfnamefont {M.~E.}\ \bibnamefont {Flatt{\'e}}},\ and\ \bibinfo {author} {\bibfnamefont {G.~D.}\ \bibnamefont {Fuchs}},\ }\bibfield  {title} {\bibinfo {title} {Temperature dependence of wavelength selectable zero-phonon emission from single defects in hexagonal boron nitride},\ }\href@noop {} {\bibfield  {journal} {\bibinfo  {journal} {Nano Letters}\ }\textbf {\bibinfo {volume} {16}},\ \bibinfo {pages} {6052} (\bibinfo {year} {2016})}\BibitemShut {NoStop}%
\bibitem [{\citenamefont {Noh}\ \emph {et~al.}(2018)\citenamefont {Noh}, \citenamefont {Choi}, \citenamefont {Kim}, \citenamefont {Im}, \citenamefont {Kim}, \citenamefont {Seo},\ and\ \citenamefont {Lee}}]{noh2018stark}%
  \BibitemOpen
  \bibfield  {author} {\bibinfo {author} {\bibfnamefont {G.}~\bibnamefont {Noh}}, \bibinfo {author} {\bibfnamefont {D.}~\bibnamefont {Choi}}, \bibinfo {author} {\bibfnamefont {J.-H.}\ \bibnamefont {Kim}}, \bibinfo {author} {\bibfnamefont {D.-G.}\ \bibnamefont {Im}}, \bibinfo {author} {\bibfnamefont {Y.-H.}\ \bibnamefont {Kim}}, \bibinfo {author} {\bibfnamefont {H.}~\bibnamefont {Seo}},\ and\ \bibinfo {author} {\bibfnamefont {J.}~\bibnamefont {Lee}},\ }\bibfield  {title} {\bibinfo {title} {Stark tuning of single-photon emitters in hexagonal boron nitride},\ }\href@noop {} {\bibfield  {journal} {\bibinfo  {journal} {Nano Lett.}\ }\textbf {\bibinfo {volume} {18}},\ \bibinfo {pages} {4710} (\bibinfo {year} {2018})}\BibitemShut {NoStop}%
\bibitem [{\citenamefont {Xia}\ \emph {et~al.}(2019)\citenamefont {Xia}, \citenamefont {Li}, \citenamefont {Kim}, \citenamefont {Bao}, \citenamefont {Gong}, \citenamefont {Yang}, \citenamefont {Wang},\ and\ \citenamefont {Zhang}}]{xia2019room}%
  \BibitemOpen
  \bibfield  {author} {\bibinfo {author} {\bibfnamefont {Y.}~\bibnamefont {Xia}}, \bibinfo {author} {\bibfnamefont {Q.}~\bibnamefont {Li}}, \bibinfo {author} {\bibfnamefont {J.}~\bibnamefont {Kim}}, \bibinfo {author} {\bibfnamefont {W.}~\bibnamefont {Bao}}, \bibinfo {author} {\bibfnamefont {C.}~\bibnamefont {Gong}}, \bibinfo {author} {\bibfnamefont {S.}~\bibnamefont {Yang}}, \bibinfo {author} {\bibfnamefont {Y.}~\bibnamefont {Wang}},\ and\ \bibinfo {author} {\bibfnamefont {X.}~\bibnamefont {Zhang}},\ }\bibfield  {title} {\bibinfo {title} {Room-temperature giant stark effect of single photon emitter in van der waals material},\ }\href@noop {} {\bibfield  {journal} {\bibinfo  {journal} {Nano Letters}\ }\textbf {\bibinfo {volume} {19}},\ \bibinfo {pages} {7100} (\bibinfo {year} {2019})}\BibitemShut {NoStop}%
\bibitem [{\citenamefont {Nikolay}\ \emph {et~al.}(2019)\citenamefont {Nikolay}, \citenamefont {Mendelson}, \citenamefont {Sadzak}, \citenamefont {B{\"o}hm}, \citenamefont {Tran}, \citenamefont {Sontheimer}, \citenamefont {Aharonovich},\ and\ \citenamefont {Benson}}]{nikolay2019very}%
  \BibitemOpen
  \bibfield  {author} {\bibinfo {author} {\bibfnamefont {N.}~\bibnamefont {Nikolay}}, \bibinfo {author} {\bibfnamefont {N.}~\bibnamefont {Mendelson}}, \bibinfo {author} {\bibfnamefont {N.}~\bibnamefont {Sadzak}}, \bibinfo {author} {\bibfnamefont {F.}~\bibnamefont {B{\"o}hm}}, \bibinfo {author} {\bibfnamefont {T.~T.}\ \bibnamefont {Tran}}, \bibinfo {author} {\bibfnamefont {B.}~\bibnamefont {Sontheimer}}, \bibinfo {author} {\bibfnamefont {I.}~\bibnamefont {Aharonovich}},\ and\ \bibinfo {author} {\bibfnamefont {O.}~\bibnamefont {Benson}},\ }\bibfield  {title} {\bibinfo {title} {Very large and reversible stark-shift tuning of single emitters in layered hexagonal boron nitride},\ }\href@noop {} {\bibfield  {journal} {\bibinfo  {journal} {Phys. Rev. Appl.}\ }\textbf {\bibinfo {volume} {11}},\ \bibinfo {pages} {041001} (\bibinfo {year} {2019})}\BibitemShut {NoStop}%
\bibitem [{\citenamefont {Jin}\ \emph {et~al.}(2009)\citenamefont {Jin}, \citenamefont {Lin}, \citenamefont {Suenaga},\ and\ \citenamefont {Iijima}}]{jin2009fabrication}%
  \BibitemOpen
  \bibfield  {author} {\bibinfo {author} {\bibfnamefont {C.}~\bibnamefont {Jin}}, \bibinfo {author} {\bibfnamefont {F.}~\bibnamefont {Lin}}, \bibinfo {author} {\bibfnamefont {K.}~\bibnamefont {Suenaga}},\ and\ \bibinfo {author} {\bibfnamefont {S.}~\bibnamefont {Iijima}},\ }\bibfield  {title} {\bibinfo {title} {Fabrication of a freestanding boron nitride single layer and its defect assignments},\ }\href@noop {} {\bibfield  {journal} {\bibinfo  {journal} {Phys. Rev. Lett.}\ }\textbf {\bibinfo {volume} {102}},\ \bibinfo {pages} {195505} (\bibinfo {year} {2009})}\BibitemShut {NoStop}%
\bibitem [{\citenamefont {Krivanek}\ \emph {et~al.}(2010)\citenamefont {Krivanek}, \citenamefont {Chisholm}, \citenamefont {Nicolosi}, \citenamefont {Pennycook}, \citenamefont {Corbin}, \citenamefont {Dellby}, \citenamefont {Murfitt}, \citenamefont {Own}, \citenamefont {Szilagyi}, \citenamefont {Oxley} \emph {et~al.}}]{krivanek2010atom}%
  \BibitemOpen
  \bibfield  {author} {\bibinfo {author} {\bibfnamefont {O.~L.}\ \bibnamefont {Krivanek}}, \bibinfo {author} {\bibfnamefont {M.~F.}\ \bibnamefont {Chisholm}}, \bibinfo {author} {\bibfnamefont {V.}~\bibnamefont {Nicolosi}}, \bibinfo {author} {\bibfnamefont {T.~J.}\ \bibnamefont {Pennycook}}, \bibinfo {author} {\bibfnamefont {G.~J.}\ \bibnamefont {Corbin}}, \bibinfo {author} {\bibfnamefont {N.}~\bibnamefont {Dellby}}, \bibinfo {author} {\bibfnamefont {M.~F.}\ \bibnamefont {Murfitt}}, \bibinfo {author} {\bibfnamefont {C.~S.}\ \bibnamefont {Own}}, \bibinfo {author} {\bibfnamefont {Z.~S.}\ \bibnamefont {Szilagyi}}, \bibinfo {author} {\bibfnamefont {M.~P.}\ \bibnamefont {Oxley}}, \emph {et~al.},\ }\bibfield  {title} {\bibinfo {title} {Atom-by-atom structural and chemical analysis by annular dark-field electron microscopy},\ }\href@noop {} {\bibfield  {journal} {\bibinfo  {journal} {Nature}\ }\textbf {\bibinfo {volume} {464}},\ \bibinfo {pages} {571} (\bibinfo {year} {2010})}\BibitemShut {NoStop}%
\bibitem [{\citenamefont {Tawfik}\ \emph {et~al.}(2017)\citenamefont {Tawfik}, \citenamefont {Ali}, \citenamefont {Fronzi}, \citenamefont {Kianinia}, \citenamefont {Tran}, \citenamefont {Stampfl}, \citenamefont {Aharonovich}, \citenamefont {Toth},\ and\ \citenamefont {Ford}}]{tawfik2017first}%
  \BibitemOpen
  \bibfield  {author} {\bibinfo {author} {\bibfnamefont {S.~A.}\ \bibnamefont {Tawfik}}, \bibinfo {author} {\bibfnamefont {S.}~\bibnamefont {Ali}}, \bibinfo {author} {\bibfnamefont {M.}~\bibnamefont {Fronzi}}, \bibinfo {author} {\bibfnamefont {M.}~\bibnamefont {Kianinia}}, \bibinfo {author} {\bibfnamefont {T.~T.}\ \bibnamefont {Tran}}, \bibinfo {author} {\bibfnamefont {C.}~\bibnamefont {Stampfl}}, \bibinfo {author} {\bibfnamefont {I.}~\bibnamefont {Aharonovich}}, \bibinfo {author} {\bibfnamefont {M.}~\bibnamefont {Toth}},\ and\ \bibinfo {author} {\bibfnamefont {M.~J.}\ \bibnamefont {Ford}},\ }\bibfield  {title} {\bibinfo {title} {First-principles investigation of quantum emission from hbn defects},\ }\href@noop {} {\bibfield  {journal} {\bibinfo  {journal} {Nanoscale}\ }\textbf {\bibinfo {volume} {9}},\ \bibinfo {pages} {13575} (\bibinfo {year} {2017})}\BibitemShut {NoStop}%
\bibitem [{\citenamefont {Sajid}\ \emph {et~al.}(2018)\citenamefont {Sajid}, \citenamefont {Reimers},\ and\ \citenamefont {Ford}}]{sajid2018defect}%
  \BibitemOpen
  \bibfield  {author} {\bibinfo {author} {\bibfnamefont {A.}~\bibnamefont {Sajid}}, \bibinfo {author} {\bibfnamefont {J.~R.}\ \bibnamefont {Reimers}},\ and\ \bibinfo {author} {\bibfnamefont {M.~J.}\ \bibnamefont {Ford}},\ }\bibfield  {title} {\bibinfo {title} {Defect states in hexagonal boron nitride: Assignments of observed properties and prediction of properties relevant to quantum computation},\ }\href@noop {} {\bibfield  {journal} {\bibinfo  {journal} {Phys. Rev. B}\ }\textbf {\bibinfo {volume} {97}},\ \bibinfo {pages} {064101} (\bibinfo {year} {2018})}\BibitemShut {NoStop}%
\bibitem [{\citenamefont {Li}\ and\ \citenamefont {Gali}(2022{\natexlab{a}})}]{li2022bistable}%
  \BibitemOpen
  \bibfield  {author} {\bibinfo {author} {\bibfnamefont {S.}~\bibnamefont {Li}}\ and\ \bibinfo {author} {\bibfnamefont {A.}~\bibnamefont {Gali}},\ }\bibfield  {title} {\bibinfo {title} {Bistable carbon-vacancy defects in h-bn},\ }\href@noop {} {\bibfield  {journal} {\bibinfo  {journal} {Front. Quantum Sci. Technol.}\ }\textbf {\bibinfo {volume} {1}},\ \bibinfo {pages} {1007756} (\bibinfo {year} {2022}{\natexlab{a}})}\BibitemShut {NoStop}%
\bibitem [{\citenamefont {Li}\ and\ \citenamefont {Gali}(2022{\natexlab{b}})}]{li2022identification}%
  \BibitemOpen
  \bibfield  {author} {\bibinfo {author} {\bibfnamefont {S.}~\bibnamefont {Li}}\ and\ \bibinfo {author} {\bibfnamefont {A.}~\bibnamefont {Gali}},\ }\bibfield  {title} {\bibinfo {title} {Identification of an oxygen defect in hexagonal boron nitride},\ }\href@noop {} {\bibfield  {journal} {\bibinfo  {journal} {J. Phys. Chem. Lett.}\ }\textbf {\bibinfo {volume} {13}},\ \bibinfo {pages} {9544} (\bibinfo {year} {2022}{\natexlab{b}})}\BibitemShut {NoStop}%
\bibitem [{\citenamefont {Jara}\ \emph {et~al.}(2021)\citenamefont {Jara}, \citenamefont {Rauch}, \citenamefont {Botti}, \citenamefont {Marques}, \citenamefont {Norambuena}, \citenamefont {Coto}, \citenamefont {Castellanos-{\'A}guila}, \citenamefont {Maze},\ and\ \citenamefont {Munoz}}]{jara2021first}%
  \BibitemOpen
  \bibfield  {author} {\bibinfo {author} {\bibfnamefont {C.}~\bibnamefont {Jara}}, \bibinfo {author} {\bibfnamefont {T.}~\bibnamefont {Rauch}}, \bibinfo {author} {\bibfnamefont {S.}~\bibnamefont {Botti}}, \bibinfo {author} {\bibfnamefont {M.~A.}\ \bibnamefont {Marques}}, \bibinfo {author} {\bibfnamefont {A.}~\bibnamefont {Norambuena}}, \bibinfo {author} {\bibfnamefont {R.}~\bibnamefont {Coto}}, \bibinfo {author} {\bibfnamefont {J.}~\bibnamefont {Castellanos-{\'A}guila}}, \bibinfo {author} {\bibfnamefont {J.~R.}\ \bibnamefont {Maze}},\ and\ \bibinfo {author} {\bibfnamefont {F.}~\bibnamefont {Munoz}},\ }\bibfield  {title} {\bibinfo {title} {First-principles identification of single photon emitters based on carbon clusters in hexagonal boron nitride},\ }\href@noop {} {\bibfield  {journal} {\bibinfo  {journal} {J. Phys. Chem. A}\ }\textbf {\bibinfo {volume} {125}},\ \bibinfo {pages} {1325} (\bibinfo {year} {2021})}\BibitemShut {NoStop}%
\bibitem [{\citenamefont {Mendelson}\ \emph {et~al.}(2021)\citenamefont {Mendelson}, \citenamefont {Chugh}, \citenamefont {Reimers}, \citenamefont {Cheng}, \citenamefont {Gottscholl}, \citenamefont {Long}, \citenamefont {Mellor}, \citenamefont {Zettl}, \citenamefont {Dyakonov}, \citenamefont {Beton}, \citenamefont {Novikov}, \citenamefont {Jagadish}, \citenamefont {Tan}, \citenamefont {Ford}, \citenamefont {Toth}, \citenamefont {Bradac},\ and\ \citenamefont {Aharonovich}}]{mendelson2021identifying}%
  \BibitemOpen
  \bibfield  {author} {\bibinfo {author} {\bibfnamefont {N.}~\bibnamefont {Mendelson}}, \bibinfo {author} {\bibfnamefont {D.}~\bibnamefont {Chugh}}, \bibinfo {author} {\bibfnamefont {J.~R.}\ \bibnamefont {Reimers}}, \bibinfo {author} {\bibfnamefont {T.~S.}\ \bibnamefont {Cheng}}, \bibinfo {author} {\bibfnamefont {A.}~\bibnamefont {Gottscholl}}, \bibinfo {author} {\bibfnamefont {H.}~\bibnamefont {Long}}, \bibinfo {author} {\bibfnamefont {C.~J.}\ \bibnamefont {Mellor}}, \bibinfo {author} {\bibfnamefont {A.}~\bibnamefont {Zettl}}, \bibinfo {author} {\bibfnamefont {V.}~\bibnamefont {Dyakonov}}, \bibinfo {author} {\bibfnamefont {P.~H.}\ \bibnamefont {Beton}}, \bibinfo {author} {\bibfnamefont {S.~V.}\ \bibnamefont {Novikov}}, \bibinfo {author} {\bibfnamefont {C.}~\bibnamefont {Jagadish}}, \bibinfo {author} {\bibfnamefont {H.~H.}\ \bibnamefont {Tan}}, \bibinfo {author} {\bibfnamefont {M.~J.}\ \bibnamefont {Ford}}, \bibinfo {author} {\bibfnamefont {M.}~\bibnamefont {Toth}}, \bibinfo {author} {\bibfnamefont
  {C.}~\bibnamefont {Bradac}},\ and\ \bibinfo {author} {\bibfnamefont {I.}~\bibnamefont {Aharonovich}},\ }\bibfield  {title} {\bibinfo {title} {Identifying carbon as the source of visible single-photon emission from hexagonal boron nitride},\ }\href@noop {} {\bibfield  {journal} {\bibinfo  {journal} {Nat. Mater.}\ }\textbf {\bibinfo {volume} {20}},\ \bibinfo {pages} {321} (\bibinfo {year} {2021})}\BibitemShut {NoStop}%
\bibitem [{\citenamefont {Chejanovsky}\ \emph {et~al.}(2021)\citenamefont {Chejanovsky}, \citenamefont {Mukherjee}, \citenamefont {Geng}, \citenamefont {Chen}, \citenamefont {Kim}, \citenamefont {Denisenko}, \citenamefont {Finkler}, \citenamefont {Taniguchi}, \citenamefont {Watanabe}, \citenamefont {Dasari}, \citenamefont {Auburger}, \citenamefont {Gali}, \citenamefont {Smet},\ and\ \citenamefont {Wrachtrup}}]{chejanovsky2021single}%
  \BibitemOpen
  \bibfield  {author} {\bibinfo {author} {\bibfnamefont {N.}~\bibnamefont {Chejanovsky}}, \bibinfo {author} {\bibfnamefont {A.}~\bibnamefont {Mukherjee}}, \bibinfo {author} {\bibfnamefont {J.}~\bibnamefont {Geng}}, \bibinfo {author} {\bibfnamefont {Y.-C.}\ \bibnamefont {Chen}}, \bibinfo {author} {\bibfnamefont {Y.}~\bibnamefont {Kim}}, \bibinfo {author} {\bibfnamefont {A.}~\bibnamefont {Denisenko}}, \bibinfo {author} {\bibfnamefont {A.}~\bibnamefont {Finkler}}, \bibinfo {author} {\bibfnamefont {T.}~\bibnamefont {Taniguchi}}, \bibinfo {author} {\bibfnamefont {K.}~\bibnamefont {Watanabe}}, \bibinfo {author} {\bibfnamefont {D.~B.~R.}\ \bibnamefont {Dasari}}, \bibinfo {author} {\bibfnamefont {P.}~\bibnamefont {Auburger}}, \bibinfo {author} {\bibfnamefont {A.}~\bibnamefont {Gali}}, \bibinfo {author} {\bibfnamefont {J.~H.}\ \bibnamefont {Smet}},\ and\ \bibinfo {author} {\bibfnamefont {J.}~\bibnamefont {Wrachtrup}},\ }\bibfield  {title} {\bibinfo {title} {Single-spin resonance in a van der waals embedded
  paramagnetic defect},\ }\href@noop {} {\bibfield  {journal} {\bibinfo  {journal} {Nat. Mater.}\ }\textbf {\bibinfo {volume} {20}},\ \bibinfo {pages} {1079} (\bibinfo {year} {2021})}\BibitemShut {NoStop}%
\bibitem [{\citenamefont {Kresse}\ and\ \citenamefont {Furthm{\"u}ller}(1996{\natexlab{a}})}]{kresse1996efficiency}%
  \BibitemOpen
  \bibfield  {author} {\bibinfo {author} {\bibfnamefont {G.}~\bibnamefont {Kresse}}\ and\ \bibinfo {author} {\bibfnamefont {J.}~\bibnamefont {Furthm{\"u}ller}},\ }\bibfield  {title} {\bibinfo {title} {Efficiency of ab-initio total energy calculations for metals and semiconductors using a plane-wave basis set},\ }\href@noop {} {\bibfield  {journal} {\bibinfo  {journal} {Comput. Mater. Sci.}\ }\textbf {\bibinfo {volume} {6}},\ \bibinfo {pages} {15} (\bibinfo {year} {1996}{\natexlab{a}})}\BibitemShut {NoStop}%
\bibitem [{\citenamefont {Kresse}\ and\ \citenamefont {Furthm{\"u}ller}(1996{\natexlab{b}})}]{kresse1996efficient}%
  \BibitemOpen
  \bibfield  {author} {\bibinfo {author} {\bibfnamefont {G.}~\bibnamefont {Kresse}}\ and\ \bibinfo {author} {\bibfnamefont {J.}~\bibnamefont {Furthm{\"u}ller}},\ }\bibfield  {title} {\bibinfo {title} {Efficient iterative schemes for ab initio total-energy calculations using a plane-wave basis set},\ }\href@noop {} {\bibfield  {journal} {\bibinfo  {journal} {Phys. Rev. B}\ }\textbf {\bibinfo {volume} {54}},\ \bibinfo {pages} {11169} (\bibinfo {year} {1996}{\natexlab{b}})}\BibitemShut {NoStop}%
\bibitem [{\citenamefont {Bl{\"o}chl}(1994)}]{blochl1994projector}%
  \BibitemOpen
  \bibfield  {author} {\bibinfo {author} {\bibfnamefont {P.~E.}\ \bibnamefont {Bl{\"o}chl}},\ }\bibfield  {title} {\bibinfo {title} {Projector augmented-wave method},\ }\href@noop {} {\bibfield  {journal} {\bibinfo  {journal} {Phys. Rev. B}\ }\textbf {\bibinfo {volume} {50}},\ \bibinfo {pages} {17953} (\bibinfo {year} {1994})}\BibitemShut {NoStop}%
\bibitem [{\citenamefont {Kresse}\ and\ \citenamefont {Joubert}(1999)}]{kresse1999ultrasoft}%
  \BibitemOpen
  \bibfield  {author} {\bibinfo {author} {\bibfnamefont {G.}~\bibnamefont {Kresse}}\ and\ \bibinfo {author} {\bibfnamefont {D.}~\bibnamefont {Joubert}},\ }\bibfield  {title} {\bibinfo {title} {From ultrasoft pseudopotentials to the projector augmented-wave method},\ }\href@noop {} {\bibfield  {journal} {\bibinfo  {journal} {Phys. Rev. B}\ }\textbf {\bibinfo {volume} {59}},\ \bibinfo {pages} {1758} (\bibinfo {year} {1999})}\BibitemShut {NoStop}%
\bibitem [{\citenamefont {Grimme}\ \emph {et~al.}(2010)\citenamefont {Grimme}, \citenamefont {Antony}, \citenamefont {Ehrlich},\ and\ \citenamefont {Krieg}}]{grimme2010consistent}%
  \BibitemOpen
  \bibfield  {author} {\bibinfo {author} {\bibfnamefont {S.}~\bibnamefont {Grimme}}, \bibinfo {author} {\bibfnamefont {J.}~\bibnamefont {Antony}}, \bibinfo {author} {\bibfnamefont {S.}~\bibnamefont {Ehrlich}},\ and\ \bibinfo {author} {\bibfnamefont {H.}~\bibnamefont {Krieg}},\ }\bibfield  {title} {\bibinfo {title} {A consistent and accurate ab initio parametrization of density functional dispersion correction (dft-d) for the 94 elements h-pu},\ }\href@noop {} {\bibfield  {journal} {\bibinfo  {journal} {J. Chem. Phys.}\ }\textbf {\bibinfo {volume} {132}},\ \bibinfo {pages} {154104} (\bibinfo {year} {2010})}\BibitemShut {NoStop}%
\bibitem [{\citenamefont {Heyd}\ \emph {et~al.}(2003)\citenamefont {Heyd}, \citenamefont {Scuseria},\ and\ \citenamefont {Ernzerhof}}]{heyd2003hybrid}%
  \BibitemOpen
  \bibfield  {author} {\bibinfo {author} {\bibfnamefont {J.}~\bibnamefont {Heyd}}, \bibinfo {author} {\bibfnamefont {G.~E.}\ \bibnamefont {Scuseria}},\ and\ \bibinfo {author} {\bibfnamefont {M.}~\bibnamefont {Ernzerhof}},\ }\bibfield  {title} {\bibinfo {title} {Hybrid functionals based on a screened coulomb potential},\ }\href@noop {} {\bibfield  {journal} {\bibinfo  {journal} {J. Chem. Phys.}\ }\textbf {\bibinfo {volume} {118}},\ \bibinfo {pages} {8207} (\bibinfo {year} {2003})}\BibitemShut {NoStop}%
\bibitem [{\citenamefont {Cassabois}\ \emph {et~al.}(2016)\citenamefont {Cassabois}, \citenamefont {Valvin},\ and\ \citenamefont {Gil}}]{cassabois2016hexagonal}%
  \BibitemOpen
  \bibfield  {author} {\bibinfo {author} {\bibfnamefont {G.}~\bibnamefont {Cassabois}}, \bibinfo {author} {\bibfnamefont {P.}~\bibnamefont {Valvin}},\ and\ \bibinfo {author} {\bibfnamefont {B.}~\bibnamefont {Gil}},\ }\bibfield  {title} {\bibinfo {title} {Hexagonal boron nitride is an indirect bandgap semiconductor},\ }\href@noop {} {\bibfield  {journal} {\bibinfo  {journal} {Nat. Photonics}\ }\textbf {\bibinfo {volume} {10}},\ \bibinfo {pages} {262} (\bibinfo {year} {2016})}\BibitemShut {NoStop}%
\bibitem [{\citenamefont {Gali}\ \emph {et~al.}(2009)\citenamefont {Gali}, \citenamefont {Janz{\'e}n}, \citenamefont {De{\'a}k}, \citenamefont {Kresse},\ and\ \citenamefont {Kaxiras}}]{gali2009theory}%
  \BibitemOpen
  \bibfield  {author} {\bibinfo {author} {\bibfnamefont {A.}~\bibnamefont {Gali}}, \bibinfo {author} {\bibfnamefont {E.}~\bibnamefont {Janz{\'e}n}}, \bibinfo {author} {\bibfnamefont {P.}~\bibnamefont {De{\'a}k}}, \bibinfo {author} {\bibfnamefont {G.}~\bibnamefont {Kresse}},\ and\ \bibinfo {author} {\bibfnamefont {E.}~\bibnamefont {Kaxiras}},\ }\bibfield  {title} {\bibinfo {title} {Theory of spin-conserving excitation of the n- v- center in diamond},\ }\href@noop {} {\bibfield  {journal} {\bibinfo  {journal} {Phys. Rev. Lett.}\ }\textbf {\bibinfo {volume} {103}},\ \bibinfo {pages} {186404} (\bibinfo {year} {2009})}\BibitemShut {NoStop}%
\bibitem [{\citenamefont {Li}\ \emph {et~al.}(2022)\citenamefont {Li}, \citenamefont {Smart},\ and\ \citenamefont {Ping}}]{li2022carbon}%
  \BibitemOpen
  \bibfield  {author} {\bibinfo {author} {\bibfnamefont {K.}~\bibnamefont {Li}}, \bibinfo {author} {\bibfnamefont {T.~J.}\ \bibnamefont {Smart}},\ and\ \bibinfo {author} {\bibfnamefont {Y.}~\bibnamefont {Ping}},\ }\bibfield  {title} {\bibinfo {title} {Carbon trimer as a 2 ev single-photon emitter candidate in hexagonal boron nitride: A first-principles study},\ }\href@noop {} {\bibfield  {journal} {\bibinfo  {journal} {Phys. Rev. Mater.}\ }\textbf {\bibinfo {volume} {6}},\ \bibinfo {pages} {L042201} (\bibinfo {year} {2022})}\BibitemShut {NoStop}%
\bibitem [{\citenamefont {Benedek}\ \emph {et~al.}(2023)\citenamefont {Benedek}, \citenamefont {Babar}, \citenamefont {Ganyecz}, \citenamefont {Szilv{\'a}si}, \citenamefont {Legeza}, \citenamefont {Barcza},\ and\ \citenamefont {Iv{\'a}dy}}]{benedek2023symmetric}%
  \BibitemOpen
  \bibfield  {author} {\bibinfo {author} {\bibfnamefont {Z.}~\bibnamefont {Benedek}}, \bibinfo {author} {\bibfnamefont {R.}~\bibnamefont {Babar}}, \bibinfo {author} {\bibfnamefont {{\'A}.}~\bibnamefont {Ganyecz}}, \bibinfo {author} {\bibfnamefont {T.}~\bibnamefont {Szilv{\'a}si}}, \bibinfo {author} {\bibfnamefont {{\"O}.}~\bibnamefont {Legeza}}, \bibinfo {author} {\bibfnamefont {G.}~\bibnamefont {Barcza}},\ and\ \bibinfo {author} {\bibfnamefont {V.}~\bibnamefont {Iv{\'a}dy}},\ }\bibfield  {title} {\bibinfo {title} {Symmetric carbon tetramers forming spin qubits in hexagonal boron nitride},\ }\href@noop {} {\bibfield  {journal} {\bibinfo  {journal} {npj Comput. Mater.}\ }\textbf {\bibinfo {volume} {9}},\ \bibinfo {pages} {187} (\bibinfo {year} {2023})}\BibitemShut {NoStop}%
\bibitem [{\citenamefont {Guo}\ \emph {et~al.}(2023)\citenamefont {Guo}, \citenamefont {Li}, \citenamefont {Liu}, \citenamefont {Yang}, \citenamefont {Zeng}, \citenamefont {Yu}, \citenamefont {Meng}, \citenamefont {Li}, \citenamefont {Wang}, \citenamefont {Xie}, \citenamefont {Ge}, \citenamefont {Wang}, \citenamefont {Li}, \citenamefont {Xu}, \citenamefont {Wang}, \citenamefont {Tang}, \citenamefont {Gali}, \citenamefont {Li},\ and\ \citenamefont {Guo}}]{guo2023coherent}%
  \BibitemOpen
  \bibfield  {author} {\bibinfo {author} {\bibfnamefont {N.-J.}\ \bibnamefont {Guo}}, \bibinfo {author} {\bibfnamefont {S.}~\bibnamefont {Li}}, \bibinfo {author} {\bibfnamefont {W.}~\bibnamefont {Liu}}, \bibinfo {author} {\bibfnamefont {Y.-Z.}\ \bibnamefont {Yang}}, \bibinfo {author} {\bibfnamefont {X.-D.}\ \bibnamefont {Zeng}}, \bibinfo {author} {\bibfnamefont {S.}~\bibnamefont {Yu}}, \bibinfo {author} {\bibfnamefont {Y.}~\bibnamefont {Meng}}, \bibinfo {author} {\bibfnamefont {Z.-P.}\ \bibnamefont {Li}}, \bibinfo {author} {\bibfnamefont {Z.-A.}\ \bibnamefont {Wang}}, \bibinfo {author} {\bibfnamefont {L.-K.}\ \bibnamefont {Xie}}, \bibinfo {author} {\bibfnamefont {R.-C.}\ \bibnamefont {Ge}}, \bibinfo {author} {\bibfnamefont {J.-F.}\ \bibnamefont {Wang}}, \bibinfo {author} {\bibfnamefont {Q.}~\bibnamefont {Li}}, \bibinfo {author} {\bibfnamefont {J.-S.}\ \bibnamefont {Xu}}, \bibinfo {author} {\bibfnamefont {Y.-T.}\ \bibnamefont {Wang}}, \bibinfo {author} {\bibfnamefont {J.-S.}\ \bibnamefont {Tang}}, \bibinfo
  {author} {\bibfnamefont {A.}~\bibnamefont {Gali}}, \bibinfo {author} {\bibfnamefont {C.-F.}\ \bibnamefont {Li}},\ and\ \bibinfo {author} {\bibfnamefont {G.-C.}\ \bibnamefont {Guo}},\ }\bibfield  {title} {\bibinfo {title} {Coherent control of an ultrabright single spin in hexagonal boron nitride at room temperature},\ }\href@noop {} {\bibfield  {journal} {\bibinfo  {journal} {Nat. Commun.}\ }\textbf {\bibinfo {volume} {14}},\ \bibinfo {pages} {2893} (\bibinfo {year} {2023})}\BibitemShut {NoStop}%
\bibitem [{\citenamefont {Turiansky}\ \emph {et~al.}(2019)\citenamefont {Turiansky}, \citenamefont {Alkauskas}, \citenamefont {Bassett},\ and\ \citenamefont {Van~de Walle}}]{turiansky2019dangling}%
  \BibitemOpen
  \bibfield  {author} {\bibinfo {author} {\bibfnamefont {M.~E.}\ \bibnamefont {Turiansky}}, \bibinfo {author} {\bibfnamefont {A.}~\bibnamefont {Alkauskas}}, \bibinfo {author} {\bibfnamefont {L.~C.}\ \bibnamefont {Bassett}},\ and\ \bibinfo {author} {\bibfnamefont {C.~G.}\ \bibnamefont {Van~de Walle}},\ }\bibfield  {title} {\bibinfo {title} {Dangling bonds in hexagonal boron nitride as single-photon emitters},\ }\href@noop {} {\bibfield  {journal} {\bibinfo  {journal} {Phys. Rev. Lett.}\ }\textbf {\bibinfo {volume} {123}},\ \bibinfo {pages} {127401} (\bibinfo {year} {2019})}\BibitemShut {NoStop}%
\bibitem [{\citenamefont {Li}\ \emph {et~al.}(2025)\citenamefont {Li}, \citenamefont {Li},\ and\ \citenamefont {Gali}}]{li2025native}%
  \BibitemOpen
  \bibfield  {author} {\bibinfo {author} {\bibfnamefont {S.}~\bibnamefont {Li}}, \bibinfo {author} {\bibfnamefont {P.}~\bibnamefont {Li}},\ and\ \bibinfo {author} {\bibfnamefont {A.}~\bibnamefont {Gali}},\ }\bibfield  {title} {\bibinfo {title} {Native antisite defects in h-bn},\ }\href@noop {} {\bibfield  {journal} {\bibinfo  {journal} {arXiv preprint arXiv:2501.01133}\ } (\bibinfo {year} {2025})}\BibitemShut {NoStop}%
\bibitem [{\citenamefont {Mackoit-Sinkevi{\v{c}}ien{\.e}}\ \emph {et~al.}(2019)\citenamefont {Mackoit-Sinkevi{\v{c}}ien{\.e}}, \citenamefont {Maciaszek}, \citenamefont {Van~de Walle},\ and\ \citenamefont {Alkauskas}}]{mackoit2019carbon}%
  \BibitemOpen
  \bibfield  {author} {\bibinfo {author} {\bibfnamefont {M.}~\bibnamefont {Mackoit-Sinkevi{\v{c}}ien{\.e}}}, \bibinfo {author} {\bibfnamefont {M.}~\bibnamefont {Maciaszek}}, \bibinfo {author} {\bibfnamefont {C.~G.}\ \bibnamefont {Van~de Walle}},\ and\ \bibinfo {author} {\bibfnamefont {A.}~\bibnamefont {Alkauskas}},\ }\bibfield  {title} {\bibinfo {title} {Carbon dimer defect as a source of the 4.1 ev luminescence in hexagonal boron nitride},\ }\href@noop {} {\bibfield  {journal} {\bibinfo  {journal} {Appl. Phys. Lett.}\ }\textbf {\bibinfo {volume} {115}},\ \bibinfo {pages} {212101} (\bibinfo {year} {2019})}\BibitemShut {NoStop}%
\bibitem [{\citenamefont {Gao}\ \emph {et~al.}(2024)\citenamefont {Gao}, \citenamefont {Vaidya}, \citenamefont {Li}, \citenamefont {Dikshit}, \citenamefont {Zhang}, \citenamefont {Ju}, \citenamefont {Shen}, \citenamefont {Jin}, \citenamefont {Ping},\ and\ \citenamefont {Li}}]{gao2024single}%
  \BibitemOpen
  \bibfield  {author} {\bibinfo {author} {\bibfnamefont {X.}~\bibnamefont {Gao}}, \bibinfo {author} {\bibfnamefont {S.}~\bibnamefont {Vaidya}}, \bibinfo {author} {\bibfnamefont {K.}~\bibnamefont {Li}}, \bibinfo {author} {\bibfnamefont {S.}~\bibnamefont {Dikshit}}, \bibinfo {author} {\bibfnamefont {S.}~\bibnamefont {Zhang}}, \bibinfo {author} {\bibfnamefont {P.}~\bibnamefont {Ju}}, \bibinfo {author} {\bibfnamefont {K.}~\bibnamefont {Shen}}, \bibinfo {author} {\bibfnamefont {Y.}~\bibnamefont {Jin}}, \bibinfo {author} {\bibfnamefont {Y.}~\bibnamefont {Ping}},\ and\ \bibinfo {author} {\bibfnamefont {T.}~\bibnamefont {Li}},\ }\bibfield  {title} {\bibinfo {title} {Single nuclear spin detection and control in a van der waals material},\ }\href@noop {} {\bibfield  {journal} {\bibinfo  {journal} {arXiv preprint arXiv:2409.01601}\ } (\bibinfo {year} {2024})}\BibitemShut {NoStop}%
\bibitem [{\citenamefont {Chou}\ and\ \citenamefont {Gali}(2017)}]{chou2017nitrogen}%
  \BibitemOpen
  \bibfield  {author} {\bibinfo {author} {\bibfnamefont {J.-P.}\ \bibnamefont {Chou}}\ and\ \bibinfo {author} {\bibfnamefont {A.}~\bibnamefont {Gali}},\ }\bibfield  {title} {\bibinfo {title} {Nitrogen-vacancy diamond sensor: novel diamond surfaces from ab initio simulations},\ }\href@noop {} {\bibfield  {journal} {\bibinfo  {journal} {MRS Commun.}\ }\textbf {\bibinfo {volume} {7}},\ \bibinfo {pages} {551} (\bibinfo {year} {2017})}\BibitemShut {NoStop}%
\bibitem [{\citenamefont {Laturia}\ \emph {et~al.}(2018)\citenamefont {Laturia}, \citenamefont {Van~de Put},\ and\ \citenamefont {Vandenberghe}}]{laturia2018dielectric}%
  \BibitemOpen
  \bibfield  {author} {\bibinfo {author} {\bibfnamefont {A.}~\bibnamefont {Laturia}}, \bibinfo {author} {\bibfnamefont {M.~L.}\ \bibnamefont {Van~de Put}},\ and\ \bibinfo {author} {\bibfnamefont {W.~G.}\ \bibnamefont {Vandenberghe}},\ }\bibfield  {title} {\bibinfo {title} {Dielectric properties of hexagonal boron nitride and transition metal dichalcogenides: from monolayer to bulk},\ }\href@noop {} {\bibfield  {journal} {\bibinfo  {journal} {npj 2D Mater. Appl.}\ }\textbf {\bibinfo {volume} {2}},\ \bibinfo {pages} {6} (\bibinfo {year} {2018})}\BibitemShut {NoStop}%
\bibitem [{\citenamefont {Alaerts}\ \emph {et~al.}(2024)\citenamefont {Alaerts}, \citenamefont {Xiong}, \citenamefont {Griffin},\ and\ \citenamefont {Hautier}}]{alaerts2024first}%
  \BibitemOpen
  \bibfield  {author} {\bibinfo {author} {\bibfnamefont {L.}~\bibnamefont {Alaerts}}, \bibinfo {author} {\bibfnamefont {Y.}~\bibnamefont {Xiong}}, \bibinfo {author} {\bibfnamefont {S.}~\bibnamefont {Griffin}},\ and\ \bibinfo {author} {\bibfnamefont {G.}~\bibnamefont {Hautier}},\ }\bibfield  {title} {\bibinfo {title} {First-principles study of the stark shift effect on the zero-phonon line of the nv center in diamond},\ }\href@noop {} {\bibfield  {journal} {\bibinfo  {journal} {Phys. Rev. Mater.}\ }\textbf {\bibinfo {volume} {8}},\ \bibinfo {pages} {106201} (\bibinfo {year} {2024})}\BibitemShut {NoStop}%
\bibitem [{\citenamefont {Bathen}\ \emph {et~al.}(2020)\citenamefont {Bathen}, \citenamefont {Vines},\ and\ \citenamefont {Coutinho}}]{bathen2020first}%
  \BibitemOpen
  \bibfield  {author} {\bibinfo {author} {\bibfnamefont {M.~E.}\ \bibnamefont {Bathen}}, \bibinfo {author} {\bibfnamefont {L.}~\bibnamefont {Vines}},\ and\ \bibinfo {author} {\bibfnamefont {J.}~\bibnamefont {Coutinho}},\ }\bibfield  {title} {\bibinfo {title} {First-principles calculations of stark shifts of electronic transitions for defects in semiconductors: the si vacancy in 4h-sic},\ }\href@noop {} {\bibfield  {journal} {\bibinfo  {journal} {J. Phys. Condens. Matter.}\ }\textbf {\bibinfo {volume} {33}},\ \bibinfo {pages} {075502} (\bibinfo {year} {2020})}\BibitemShut {NoStop}%
\bibitem [{\citenamefont {De~Santis}\ \emph {et~al.}(2021)\citenamefont {De~Santis}, \citenamefont {Trusheim}, \citenamefont {Chen},\ and\ \citenamefont {Englund}}]{de2021investigation}%
  \BibitemOpen
  \bibfield  {author} {\bibinfo {author} {\bibfnamefont {L.}~\bibnamefont {De~Santis}}, \bibinfo {author} {\bibfnamefont {M.~E.}\ \bibnamefont {Trusheim}}, \bibinfo {author} {\bibfnamefont {K.~C.}\ \bibnamefont {Chen}},\ and\ \bibinfo {author} {\bibfnamefont {D.~R.}\ \bibnamefont {Englund}},\ }\bibfield  {title} {\bibinfo {title} {Investigation of the stark effect on a centrosymmetric quantum emitter in diamond},\ }\href@noop {} {\bibfield  {journal} {\bibinfo  {journal} {Phys. Rev. Lett.}\ }\textbf {\bibinfo {volume} {127}},\ \bibinfo {pages} {147402} (\bibinfo {year} {2021})}\BibitemShut {NoStop}%
\bibitem [{\citenamefont {Maze}\ \emph {et~al.}(2011)\citenamefont {Maze}, \citenamefont {Gali}, \citenamefont {Togan}, \citenamefont {Chu}, \citenamefont {Trifonov}, \citenamefont {Kaxiras},\ and\ \citenamefont {Lukin}}]{maze2011properties}%
  \BibitemOpen
  \bibfield  {author} {\bibinfo {author} {\bibfnamefont {J.~R.}\ \bibnamefont {Maze}}, \bibinfo {author} {\bibfnamefont {A.}~\bibnamefont {Gali}}, \bibinfo {author} {\bibfnamefont {E.}~\bibnamefont {Togan}}, \bibinfo {author} {\bibfnamefont {Y.}~\bibnamefont {Chu}}, \bibinfo {author} {\bibfnamefont {A.}~\bibnamefont {Trifonov}}, \bibinfo {author} {\bibfnamefont {E.}~\bibnamefont {Kaxiras}},\ and\ \bibinfo {author} {\bibfnamefont {M.~D.}\ \bibnamefont {Lukin}},\ }\bibfield  {title} {\bibinfo {title} {Properties of nitrogen-vacancy centers in diamond: the group theoretic approach},\ }\href@noop {} {\bibfield  {journal} {\bibinfo  {journal} {New J. Phys.}\ }\textbf {\bibinfo {volume} {13}},\ \bibinfo {pages} {025025} (\bibinfo {year} {2011})}\BibitemShut {NoStop}%
\bibitem [{\citenamefont {Udvarhelyi}\ \emph {et~al.}(2023)\citenamefont {Udvarhelyi}, \citenamefont {Clua-Provost}, \citenamefont {Durand}, \citenamefont {Li}, \citenamefont {Edgar}, \citenamefont {Gil}, \citenamefont {Cassabois}, \citenamefont {Jacques},\ and\ \citenamefont {Gali}}]{udvarhelyi2023planar}%
  \BibitemOpen
  \bibfield  {author} {\bibinfo {author} {\bibfnamefont {P.}~\bibnamefont {Udvarhelyi}}, \bibinfo {author} {\bibfnamefont {T.}~\bibnamefont {Clua-Provost}}, \bibinfo {author} {\bibfnamefont {A.}~\bibnamefont {Durand}}, \bibinfo {author} {\bibfnamefont {J.}~\bibnamefont {Li}}, \bibinfo {author} {\bibfnamefont {J.~H.}\ \bibnamefont {Edgar}}, \bibinfo {author} {\bibfnamefont {B.}~\bibnamefont {Gil}}, \bibinfo {author} {\bibfnamefont {G.}~\bibnamefont {Cassabois}}, \bibinfo {author} {\bibfnamefont {V.}~\bibnamefont {Jacques}},\ and\ \bibinfo {author} {\bibfnamefont {A.}~\bibnamefont {Gali}},\ }\bibfield  {title} {\bibinfo {title} {A planar defect spin sensor in a two-dimensional material susceptible to strain and electric fields},\ }\href@noop {} {\bibfield  {journal} {\bibinfo  {journal} {npj Comput. Mater.}\ }\textbf {\bibinfo {volume} {9}},\ \bibinfo {pages} {150} (\bibinfo {year} {2023})}\BibitemShut {NoStop}%
\bibitem [{\citenamefont {Iwa{\'n}ski}\ \emph {et~al.}(2024)\citenamefont {Iwa{\'n}ski}, \citenamefont {Korona}, \citenamefont {Tokarczyk}, \citenamefont {Kowalski}, \citenamefont {D{\k{a}}browska}, \citenamefont {Tatarczak}, \citenamefont {Rogala}, \citenamefont {Bilska}, \citenamefont {W{\'o}jcik}, \citenamefont {Kret} \emph {et~al.}}]{iwanski2024revealing}%
  \BibitemOpen
  \bibfield  {author} {\bibinfo {author} {\bibfnamefont {J.}~\bibnamefont {Iwa{\'n}ski}}, \bibinfo {author} {\bibfnamefont {K.~P.}\ \bibnamefont {Korona}}, \bibinfo {author} {\bibfnamefont {M.}~\bibnamefont {Tokarczyk}}, \bibinfo {author} {\bibfnamefont {G.}~\bibnamefont {Kowalski}}, \bibinfo {author} {\bibfnamefont {A.~K.}\ \bibnamefont {D{\k{a}}browska}}, \bibinfo {author} {\bibfnamefont {P.}~\bibnamefont {Tatarczak}}, \bibinfo {author} {\bibfnamefont {I.}~\bibnamefont {Rogala}}, \bibinfo {author} {\bibfnamefont {M.}~\bibnamefont {Bilska}}, \bibinfo {author} {\bibfnamefont {M.}~\bibnamefont {W{\'o}jcik}}, \bibinfo {author} {\bibfnamefont {S.}~\bibnamefont {Kret}}, \emph {et~al.},\ }\bibfield  {title} {\bibinfo {title} {Revealing polytypism in 2d boron nitride with uv photoluminescence},\ }\href@noop {} {\bibfield  {journal} {\bibinfo  {journal} {npj 2D Mater. Appl.}\ }\textbf {\bibinfo {volume} {8}},\ \bibinfo {pages} {72} (\bibinfo {year} {2024})}\BibitemShut {NoStop}%
\bibitem [{\citenamefont {Ohba}\ \emph {et~al.}(2001)\citenamefont {Ohba}, \citenamefont {Miwa}, \citenamefont {Nagasako},\ and\ \citenamefont {Fukumoto}}]{ohba2001first}%
  \BibitemOpen
  \bibfield  {author} {\bibinfo {author} {\bibfnamefont {N.}~\bibnamefont {Ohba}}, \bibinfo {author} {\bibfnamefont {K.}~\bibnamefont {Miwa}}, \bibinfo {author} {\bibfnamefont {N.}~\bibnamefont {Nagasako}},\ and\ \bibinfo {author} {\bibfnamefont {A.}~\bibnamefont {Fukumoto}},\ }\bibfield  {title} {\bibinfo {title} {First-principles study on structural, dielectric, and dynamical properties for three bn polytypes},\ }\href@noop {} {\bibfield  {journal} {\bibinfo  {journal} {Phys. Rev. B}\ }\textbf {\bibinfo {volume} {63}},\ \bibinfo {pages} {115207} (\bibinfo {year} {2001})}\BibitemShut {NoStop}%
\bibitem [{\citenamefont {Kumar}\ \emph {et~al.}(2016)\citenamefont {Kumar}, \citenamefont {Chauhan}, \citenamefont {Agarwal},\ and\ \citenamefont {Bhowmick}}]{kumar2016thickness}%
  \BibitemOpen
  \bibfield  {author} {\bibinfo {author} {\bibfnamefont {P.}~\bibnamefont {Kumar}}, \bibinfo {author} {\bibfnamefont {Y.~S.}\ \bibnamefont {Chauhan}}, \bibinfo {author} {\bibfnamefont {A.}~\bibnamefont {Agarwal}},\ and\ \bibinfo {author} {\bibfnamefont {S.}~\bibnamefont {Bhowmick}},\ }\bibfield  {title} {\bibinfo {title} {Thickness and stacking dependent polarizability and dielectric constant of graphene--hexagonal boron nitride composite stacks},\ }\href@noop {} {\bibfield  {journal} {\bibinfo  {journal} {J. Phys. Chem. C}\ }\textbf {\bibinfo {volume} {120}},\ \bibinfo {pages} {17620} (\bibinfo {year} {2016})}\BibitemShut {NoStop}%
\bibitem [{\citenamefont {Zhigulin}\ \emph {et~al.}(2023)\citenamefont {Zhigulin}, \citenamefont {Horder}, \citenamefont {Iv{\'a}dy}, \citenamefont {White}, \citenamefont {Gale}, \citenamefont {Li}, \citenamefont {Lobo}, \citenamefont {Toth}, \citenamefont {Aharonovich},\ and\ \citenamefont {Kianinia}}]{zhigulin2023stark}%
  \BibitemOpen
  \bibfield  {author} {\bibinfo {author} {\bibfnamefont {I.}~\bibnamefont {Zhigulin}}, \bibinfo {author} {\bibfnamefont {J.}~\bibnamefont {Horder}}, \bibinfo {author} {\bibfnamefont {V.}~\bibnamefont {Iv{\'a}dy}}, \bibinfo {author} {\bibfnamefont {S.~J.}\ \bibnamefont {White}}, \bibinfo {author} {\bibfnamefont {A.}~\bibnamefont {Gale}}, \bibinfo {author} {\bibfnamefont {C.}~\bibnamefont {Li}}, \bibinfo {author} {\bibfnamefont {C.~J.}\ \bibnamefont {Lobo}}, \bibinfo {author} {\bibfnamefont {M.}~\bibnamefont {Toth}}, \bibinfo {author} {\bibfnamefont {I.}~\bibnamefont {Aharonovich}},\ and\ \bibinfo {author} {\bibfnamefont {M.}~\bibnamefont {Kianinia}},\ }\bibfield  {title} {\bibinfo {title} {Stark effect of blue quantum emitters in hexagonal boron nitride},\ }\href@noop {} {\bibfield  {journal} {\bibinfo  {journal} {Phys. Rev. Appl.}\ }\textbf {\bibinfo {volume} {19}},\ \bibinfo {pages} {044011} (\bibinfo {year} {2023})}\BibitemShut {NoStop}%
\bibitem [{\citenamefont {King-Smith}\ and\ \citenamefont {Vanderbilt}(1993)}]{king1993theory}%
  \BibitemOpen
  \bibfield  {author} {\bibinfo {author} {\bibfnamefont {R.}~\bibnamefont {King-Smith}}\ and\ \bibinfo {author} {\bibfnamefont {D.}~\bibnamefont {Vanderbilt}},\ }\bibfield  {title} {\bibinfo {title} {Theory of polarization of crystalline solids},\ }\href@noop {} {\bibfield  {journal} {\bibinfo  {journal} {Phys. Rev. B}\ }\textbf {\bibinfo {volume} {47}},\ \bibinfo {pages} {1651} (\bibinfo {year} {1993})}\BibitemShut {NoStop}%
\bibitem [{\citenamefont {Resta}(1994)}]{resta1994macroscopic}%
  \BibitemOpen
  \bibfield  {author} {\bibinfo {author} {\bibfnamefont {R.}~\bibnamefont {Resta}},\ }\bibfield  {title} {\bibinfo {title} {Macroscopic polarization in crystalline dielectrics: the geometric phase approach},\ }\href@noop {} {\bibfield  {journal} {\bibinfo  {journal} {Rev. Mod. Phys.}\ }\textbf {\bibinfo {volume} {66}},\ \bibinfo {pages} {899} (\bibinfo {year} {1994})}\BibitemShut {NoStop}%
\bibitem [{\citenamefont {Spaldin}(2012)}]{spaldin2012beginner}%
  \BibitemOpen
  \bibfield  {author} {\bibinfo {author} {\bibfnamefont {N.~A.}\ \bibnamefont {Spaldin}},\ }\bibfield  {title} {\bibinfo {title} {A beginner's guide to the modern theory of polarization},\ }\href@noop {} {\bibfield  {journal} {\bibinfo  {journal} {J. Solid State Chem.}\ }\textbf {\bibinfo {volume} {195}},\ \bibinfo {pages} {2} (\bibinfo {year} {2012})}\BibitemShut {NoStop}%
\bibitem [{\citenamefont {Udvarhelyi}\ \emph {et~al.}(2019)\citenamefont {Udvarhelyi}, \citenamefont {Nagy}, \citenamefont {Kaiser}, \citenamefont {Lee}, \citenamefont {Wrachtrup},\ and\ \citenamefont {Gali}}]{udvarhelyi2019spectrally}%
  \BibitemOpen
  \bibfield  {author} {\bibinfo {author} {\bibfnamefont {P.}~\bibnamefont {Udvarhelyi}}, \bibinfo {author} {\bibfnamefont {R.}~\bibnamefont {Nagy}}, \bibinfo {author} {\bibfnamefont {F.}~\bibnamefont {Kaiser}}, \bibinfo {author} {\bibfnamefont {S.-Y.}\ \bibnamefont {Lee}}, \bibinfo {author} {\bibfnamefont {J.}~\bibnamefont {Wrachtrup}},\ and\ \bibinfo {author} {\bibfnamefont {A.}~\bibnamefont {Gali}},\ }\bibfield  {title} {\bibinfo {title} {Spectrally stable defect qubits with no inversion symmetry for robust spin-to-photon interface},\ }\href@noop {} {\bibfield  {journal} {\bibinfo  {journal} {Phys. Rev. Appl.}\ }\textbf {\bibinfo {volume} {11}},\ \bibinfo {pages} {044022} (\bibinfo {year} {2019})}\BibitemShut {NoStop}%
\end{thebibliography}%

\end{document}